%% file: strong.tex
\magnification=\magstephalf
\input header

\input gaugedefs

\input spinordef

\input epsf
\def\captionsize{\ninerm}

\SetUpAuxFile
\hfuzz 20pt
\overfullrule 0pt

\def\e{\epsilon}
\def\tree{{\rm tree\vphantom{p}}}
\def\dash{\hbox{-\kern-.02em}}
\def\oneloop{{\rm 1\dash{}loop}}

\def\lloop{\dash{}{\rm loop}}
\def\dash{\hbox{-\kern-.02em}}
\def\Split{\mathop{\rm C}\nolimits}
\def\Soft{\mathop{\rm Soft}\nolimits}

\def\ah{{\hat a}}
\def\bh{{\smash{{\hat b}}{}}}
\def\Ant{\mathop{\rm Ant}\nolimits}
\def\Ctree{\Split^\tree}

\def\LIPS{{\rm LIPS}}
\def\phpol{{\rm ph.\ pol.\ }}
\def\llongrightarrow{%
\relbar\mskip-0.5mu\joinrel\mskip-0.5mu\relbar\mskip-0.5mu\joinrel\longrightarrow}
\def\inlimit^#1{\buildrel#1\over\llongrightarrow}
\def\frac#1#2{{#1\over #2}}
\def\sing#1{\lfloor #1\rceil}
\def\lu#1{{\lambda_{#1}}}
\def\la{\lu{a}}
\def\lb{\lu{b}}

\def\ls{\lu{s}}

\preref\DixonTASI{L.\ Dixon, in 
{\it QCD \& Beyond: Proceedings of TASI '95}, 
ed. D.\ E.\ Soper (World Scientific, 1996) [hep-ph/9601359]}
\preref\Recurrence{F.\ A.\ Berends and W.\ T.\ Giele, 
Nucl.\ Phys.\ B306:759 (1988)}
\preref\HelicityRecurrence{
D.\ A.\ Kosower,
Nucl.\ Phys.\ B335:23 (1990)%
}
\preref\BernChalmers{
Z.\ Bern and G.\ Chalmers,
Nucl.\ Phys.\ B447:465 (1995) [hep-ph/9503236]%
}
\preref\ManganoParke{M.\ Mangano and S.\ J.\ Parke, Phys.\ Rep.\ 200:301 (1991)}
\preref\SingleAntenna{
D.\ A.\ Kosower,
Phys.\ Rev.\ D57:5410 (1998) [hep-ph/9710213]%
}
\preref\MultipleAntenna{
D.\ A.\ Kosower,
Phys.\ Rev.\ D67:116003 (2003) [hep-ph/0212097]%
}
\preref\SusyFour{Z. Bern, L. Dixon, D. C. Dunbar, and D. A. Kosower,
Nucl.\ Phys.\ B425:217 (1994) [hep-ph/9403226]}
\preref\AllOrdersCollinear{
D. A. Kosower,
Nucl.\ Phys.\ B552:319 (1999) [hep-ph/9901201]%
}
\preref\DDFM{
V.\ Del Duca, A.\ Frizzo and F.\ Maltoni,
Nucl.\ Phys.\ B568:211 (2000) [hep-ph/9909464]%
}
\preref\GloverCampbell{
J.\ M.\ Campbell and E.\ W.\ N.\ Glover,
Nucl.\ Phys.\ B527:264 (1998) [hep-ph/9710255]%
}
\preref\CataniGrazzini{
S.\ Catani and M.\ Grazzini,
Phys.\ Lett.\ B446:143 (1999) [hep-ph/9810389]\semi
S.\ Catani and M.\ Grazzini,
Nucl.\ Phys.\ B570:287 (2000) [hep-ph/9908523]%
}
\preref\BerendsGieleSoft{
F.\ A.\ Berends and W.\ T.\ Giele,
Nucl.\ Phys.\ B313:595 (1989)%
}
\preref\AltarelliParisi{G.\ Altarelli and G.\ Parisi, Nucl.\ Phys.\ B126:298 (1977)}
\preref\GieleGlover{W.\ T.\ Giele and E.\ W.\ N.\ Glover, 
Phys.\ Rev.\ D46:1980 (1992)}
\preref\GieleGloverKosower{
W.\ T.\ Giele, E.\ W.\ N.\ Glover and D.\ A.\ Kosower,
Nucl.\ Phys.\ B403:633 (1993) [hep-ph/9302225]%
}
\preref\CataniSeymour{
S.\ Catani and M.\ H.\ Seymour,
Phys.\ Lett.\ B378:287 (1996) [hep-ph/9602277]\semi
S.\ Catani and M.\ H.\ Seymour,
Nucl.\ Phys.\ B485:291 (1997); erratum-ibid.\ B510:503 (1997) [hep-ph/9605323]%
}
\preref\Color{%
F.\ A.\ Berends and W.\ T.\ Giele,
Nucl.\ Phys.\ B294:700 (1987)\semi
D.\ A.\ Kosower, B.-H.\ Lee and V.\ P.\ Nair, Phys.\ Lett.\ 201B:85 (1988)\semi
M.\ Mangano, S.\ Parke and Z.\ Xu, Nucl.\ Phys.\ B298:653 (1988)\semi
Z.\ Bern and D.\ A.\ Kosower, Nucl.\ Phys.\ B362:389 (1991)}
\preref\qqggg{Z. Bern, L. Dixon, and D. A. Kosower,
Nucl.\ Phys.\  B437:259 (1995) [hep-ph/9409393]}
\preref\AlternateColorDecomposition{
V.\ Del Duca, L.\ J.\ Dixon and F.\ Maltoni,
Nucl.\ Phys.\ B571: 51 (2000) [hep-ph/9910563]%
}
\preref\UnitarityB{%
Z.\ Bern, L.\ J.\ Dixon, D.\ C.\ Dunbar and D.\ A.\ Kosower,
Nucl.\ Phys.\ B435:59 (1995) [hep-ph/9409265].
}
\preref\UnitarityReview{
Z.\ Bern, L.\ Dixon, and D.\ A.\ Kosower,
Ann.\ Rev.\ Nucl.\ Part.\ Sci.\ 46:109 (1996) [hep-ph/9602280]}
\preref\HigherLoopAntenna{
D.\ A.\ Kosower,
hep-ph/0301069%
}
\preref\BernMorgan{
Z.\ Bern and A.\ G.\ Morgan,
Nucl.\ Phys.\ B467:479 (1996) [hep-ph/9511336]%
}

\preref\CataniConjecture{
S.\ Catani,
Phys.\ Lett.\ B427:161 (1998) [hep-ph/9802439]%
}
\preref\StermanTejeda{
G.\ Sterman and M.\ E.\ Tejeda-Yeomans,
Phys.\ Lett.\ B552:48 (2003) [hep-ph/0210130]%
}
\preref\vanNeerven{W.\ L.\ van\ Neerven, \NPB 268:453 (1986)}
\preref\CataniGrazziniSoft{
S.\ Catani and M.\ Grazzini,
Nucl.\ Phys.\ B591:435 (2000) [hep-ph/0007142]%
}
\preref\FKS{
S.\ Frixione, Z.\ Kunszt and A.\ Signer,
Nucl.\ Phys.\ B467:399 (1996) [hep-ph/9512328]
}
\preref\Byckling{E. Byckling and K. Kajantie, {\it Particle Kinematics\/}
(Wiley, 1973)}
\preref\OneloopSplitB{
Z.\ Bern, V.\ Del Duca and C.\ R.\ Schmidt,
Phys.\ Lett.\ B445:168 (1998) [hep-ph/9810409]\semi
Z.\ Bern, V.\ Del Duca, W.\ B.\ Kilgore and C.\ R.\ Schmidt,
Phys.\ Rev.\ D60:116001 (1999) [hep-ph/9903516]
}
\preref\OneloopSplitA{
D.\ A.\ Kosower and P.\ Uwer,
Nucl.\ Phys.\ B563:477 (1999) [hep-ph/9903515]
}

\loadfourteenpoint
\noindent\nopagenumbers
[hep-ph/0311272] \hfill{Saclay/SPhT--T03/178}

\leftlabelstrue
\vskip -0.7 in
\Title{Antenna Factorization in Strongly-Ordered Limits}
\vskip 10pt

\baselineskip17truept
\centerline{David A. Kosower}
\baselineskip12truept
\centerline{\it Service de Physique Th\'eorique${}^{\natural}$}
\centerline{\it Centre d'Etudes de Saclay}
\centerline{\it F-91191 Gif-sur-Yvette cedex, France}
\centerline{\tt kosower@spht.saclay.cea.fr}

\vskip 0.2in\baselineskip13truept

\vskip 0.5truein
\centerline{\bf Abstract}
{\narrower 

When energies or angles of gluons emitted in a gauge-theory process
are small and strongly ordered,
the emission factorizes in a simple way to all orders
in perturbation theory.  I show how to unify the various strongly-ordered
soft, mixed soft-collinear, and collinear limits using antenna factorization
amplitudes, which are generalizations of the 
Catani--Seymour dipole factorization function.

}
\vskip 0.3truein


\vfill
\vskip 0.1in
\noindent\hrule width 3.6in\hfil\break
\noindent
${}^{\natural}$Laboratory of the
{\it Direction des Sciences de la Mati\`ere\/}
of the {\it Commissariat \`a l'Energie Atomique\/} of France.\hfil\break

\Date{}

\line{}

\baselineskip17pt
%

\section{Introduction}
\vskip 10pt

The properties of gauge-theory amplitudes in degenerate limits 
play an important role in the formalism of perturbative QCD.  
In addition to the usual ultraviolet singularities present when 
amplitudes are expressed in terms of a bare coupling, 
intermediate quantities also contain infrared divergences.
All these divergences may be regulated by a
dimensional regulator $\e=(4-D)/2$.  Infrared divergences arise
both from loop integrations and from the integration over singular
regions of phase space.  Both have a universal structure, which
allows the ultimate cancellation of these divergences to be treated
in a universal manner as 
well~[\use\GieleGlover,\use\GieleGloverKosower,\use\FKS,\use\CataniSeymour].
The universality of real-emission infrared divergences emerges from
the universality of gauge-theory amplitudes in soft and collinear
limits of external momenta.  This universality is reflected in the
factorization~[\use\ManganoParke,\use\SusyFour] 
of amplitudes into `hard' parts, independent of the
soft or collinear emission, and emission amplitudes for soft or collinear partons,
whose singular behavior is independent of the details of the `hard'
process.

Traditionally, the soft and collinear limits were treated independently;
but in computing phase-space integrals, this requires handling the boundary
in between them.  It is nicer, and perhaps indispensable beyond next-to-leading
order, to combine these two limits.  This was done by Catani and 
Seymour~[\use\CataniSeymour] for single emission at the 
level of the amplitude squared.  It can also be
carried out at the amplitude level, through the definition of {\it antenna\/}
factorization amplitudes~[\use\SingleAntenna].  The latter construct generalizes
nicely to the emission of multiple soft or collinear radiation~[\use\MultipleAntenna].

When examining the emission of two soft gluons in a process, there are
two distinct regimes we can consider: that in which the gluon
energies are comparable, and that in which they are strongly ordered, $E_1\ll E_2$.
The latter generalizes to the emission of $n$ soft gluons,
$E_1\ll E_2 \ll \cdots \ll E_n$.
This strongly-ordered limit is of interest because the leading
singularities arise there --- the leading powers of $\e^{-1}$ in intermediate quantities
as well as the leading large logarithms in ultimate physical quantities.

In QED, soft photon emission not only factorizes, but factorizes
independently of other photons.  Multiple soft photon emission thus
follows straightforwardly from single photon emission, and there is no
substantive distinction between unordered and strongly-ordered
emission.  In non-Abelian gauge theories, the situation is more
complicated.  Berends and Giele presented~[\use\BerendsGieleSoft] a
general form for strongly-ordered multiple soft emission.  The purpose of the present
paper is to generalize their discussion to include strongly-ordered mixed
soft-collinear and multiply-collinear emission as well.

  The properties of non-Abelian 
gauge-theory amplitudes in singular limits are easiest to understand
in the context of a color decomposition~[\use\Color].  In the present
paper, I will concentrate on all-gluon amplitudes, though the formalism
readily extends to amplitudes with quarks and (colored) scalars as well. For 
tree-level all-gluon amplitudes in an $SU(N)$ gauge theory 
the color decomposition has the form,
$$
{\cal A}_n^\tree(\{k_i,\lambda_i,a_i\}) = 
\sum_{\sigma \in S_n/Z_n} \Tr(T^{a_{\sigma(1)}}\cdots T^{a_{\sigma(n)}})\,
A_n^\tree(\sigma(1^{\lambda_1},\ldots,n^{\lambda_n}))\,,
\eqn\TreeColorDecomposition$$
where $S_n/Z_n$ is the group of non-cyclic permutations
on $n$ symbols, and $j^{\lambda_j}$ denotes the $j$-th momentum
and helicity $\lambda_j$.  The notation $j_1+j_2$ appearing below will denote
the sum of momenta, $k_{j_1}+k_{j_2}$.
I use the normalization $\Tr(T^a T^b) = \delta^{ab}$.
Analogous formul\ae\ hold for amplitudes
with quark-antiquark pairs or uncolored external lines.
The color-ordered or partial amplitude $A_n$ is gauge invariant, and has
simple factorization properties in both the soft and collinear limits,
$$
\eqalign{
A_{n}^\tree(\ldots,a,s^{\lambda_s},b,\ldots)
 &\inlimit^{k_s\rightarrow 0}
  \Soft^\tree(a,s^{\lambda_s},b)\,
      A_{n-1}^\tree(\ldots,a,b,\ldots),\cr
A_{n}^\tree(\ldots,a^{\lambda_a},b^{\lambda_b},\ldots)
 &\inlimit^{a \parallel b}
\sum_{\lambda=\pm}  
  \Split^{\rm tree}_{-\lambda}(a^{\lambda_a},b^{\lambda_b};z)\,
      A_{n-1}^\tree(\ldots,(a+b)^\lambda,\ldots).\cr
}\eqn\Factorization
$$

The collinear splitting amplitude $\Split^{\rm tree}$, 
squared and summed over helicities,
gives the usual unpolarized Altarelli--Parisi splitting 
function~[\use\AltarelliParisi].
It depends on the collinear momentum fraction $z$ (here made 
explicit) in addition to invariants built out of the collinear momenta.
While the complete amplitude also factorizes in the collinear limit,
the same is not true of the soft limit; the eikonal factors $\Soft^\tree$
get tangled up with the color structure via the sum~(\use\TreeColorDecomposition). 
It is for this reason that the color decomposition is useful.  

I review factorization in strongly-ordered soft limits
in section~\use\MultipleSoftSection{}, and consider a subtlety with strongly-ordered
collinear limits in section~\use\MultipleCollinearSection.  
The soft and collinear limits can be unified using the antenna factorization
amplitude, which I review in section~\use\AntennaReviewSection.  The
form of the resulting
factorization is similar to that for the soft factorization in 
eqn.~(\use\Factorization).  
I also review the tree-level
multiple-emission antenna amplitude, and show in section~\use\IteratedAntennaSection{}
how it can be simplified
in strongly-ordered limits.  These simplifications generalize readily
beyond tree amplitudes, as discussed
in section~\use\LoopAntennaSection.  

\def\tp{\!+\!}\def\tm{\!-\!}

\section{Multiple Soft Emission}
\tagsection\MultipleSoftSection
\vskip 10pt

Color-ordered amplitudes have simple factorization properties in limits where
several legs become soft simultaneously.  These are trivial generalizations of
eqn.~(\use\Factorization) if the soft legs are not color-connected, that is are
not neighboring arguments to the amplitude,
$$\eqalign{
A_{n}^\tree(\ldots,a_1,&s_1^{\lambda_{s_1}},b_1,\ldots,a_2,s_2^{\lambda_{s_2}},b_2,\ldots)
 \inlimit^{k_{s_1},k_{s_2}\rightarrow 0}
\cr &
  \Soft^\tree(a_1,s_1^{\lambda_{s_1}},b_1)\,
  \Soft^\tree(a_2,s_2^{\lambda_{s_2}},b_2)\,
      A_{n-2}^\tree(\ldots,a_1,b_1,\ldots,a_2,b_2,\ldots)\,.
}\anoneqn$$
When the two soft legs are color-connected, in general there is no such simple
decomposition of the soft factor itself, and we have~[\use\BerendsGieleSoft]
$$
A_{n}^\tree(\ldots,a,s_1^{\lambda_{s_1}},s_2^{\lambda_{s_2}},b,\ldots)
 \inlimit^{k_{s_1},k_{s_2}\rightarrow 0}
  \Soft^\tree(a,s_1^{\lambda_{s_1}},s_2^{\lambda_{s_2}},b)\,
      A_{n-2}^\tree(\ldots,a,b,\ldots)\,.
\anoneqn$$
However, in the limit where one of the gluons is much softer than the other --- say 
$k_{s_1}\ll k_{s_2}$, the soft factor itself factorizes,
$$
  \Soft^\tree(a,s_1^{\lambda_{s_1}},s_2^{\lambda_{s_2}},b)\,
 \inlimit^{k_{s_1} \ll k_{s_2}}
  \Soft^\tree(a,s_1^{\lambda_{s_1}},s_2)\,
  \Soft^\tree(a,s_2^{\lambda_{s_2}},b).
\anoneqn$$
The soft factors are nested or iterated, a feature of the factorization that
will have an echo in the strongly-ordered antenna factorization to be discussed
in section~\use\IteratedAntennaSection.

Similar results hold for multiple-gluon emission; the amplitude factorizes,
$$
A_{n}^\tree(\ldots,a,s_1^{\lambda_{s_1}},\ldots,s_m^{\lambda_{s_m}},b,\ldots)
 \inlimit^{k_{s_1},\ldots,k_{s_m}\rightarrow 0}
  \Soft^\tree(a,s_1^{\lambda_{s_1}},\ldots,s_m^{\lambda_{s_m}},b)\,
      A_{n-m}^\tree(\ldots,a,b,\ldots),
\anoneqn$$
and in the strongly-ordered domain, the soft factor itself factorizes,
$$\eqalign{
  &\Soft^\tree(a,s_1^{\lambda_{s_1}},s_2^{\lambda_{s_2}},\ldots,
                s_m^{\lambda_{s_m}},b)\,
 \inlimit^{k_{s_1} \ll k_{s_2} \ll \cdots \ll k_{s_m}}\cr
&\hskip 20mm
  \Soft^\tree(a,s_1^{\lambda_{s_1}},s_2)\,
  \Soft^\tree(a,s_2^{\lambda_{s_2}},s_3)\cdots
  \Soft^\tree(a,s_m^{\lambda_{s_m}},b) + {\rm subleading}.
}\anoneqn$$

Because the soft factors are independent of the helicities of the hard legs
(they depend only on the helicities of the soft gluons themselves), these 
strong-ordering simplifications square in a simple way even after summing over
final helicities and averaging over initial ones,
$$\eqalign{
  &\bigl\langle|\Soft^\tree(a,s_1^{\lambda_{s_1}},s_2^{\lambda_{s_2}},\ldots,
                s_m^{\lambda_{s_m}},b)|^2\bigr\rangle
 \inlimit^{k_{s_1} \ll k_{s_2} \ll \cdots \ll k_{s_m}}\cr
&\hskip 15mm
  \bigl\langle|\Soft^\tree(a,s_1^{\lambda_{s_1}},s_2)|^2\bigr\rangle
  \bigl\langle|\Soft^\tree(a,s_2^{\lambda_{s_2}},s_3)|^2\bigr\rangle\cdots
  \bigl\langle|\Soft^\tree(a,s_m^{\lambda_{s_m}},b)|^2\bigr\rangle
 +{\rm subleading},
}\anoneqn$$
where $\langle\rangle$ denotes helicity summation and averaging.
This simplification carries over directly to the structure of the leading-color
term in the squared matrix element, which contains no interference terms between
different permutations of arguments to the color-ordered amplitudes.

\section{Iterated Collinear Limits and Azimuthal Averaging}
\tagsection\MultipleCollinearSection
\vskip 10pt

In limits where several neighboring legs become collinear, the color-ordered
amplitude again factorizes,
$$
\eqalign{
&A_{n}^\tree(\ldots,a^{\lambda_a},b^{\lambda_b},\ldots)
 \inlimit^{a_1 \parallel a_2\parallel \cdots \parallel a_m}
\cr &\hskip 20mm
\sum_{\lambda=\pm}  
  \Split^{\rm tree}_{-\lambda}(a_1^{\lambda_{a_1}},a_2^{\lambda_{a_2}},\ldots,
                               a_m^{\lambda_{a_m}};\{z_i\})\,
      A_{n-m}^\tree(\ldots,(a_1+\cdots +a_m)^\lambda,\ldots).
}\anoneqn$$
The $z_i$ denote the momentum fractions of the collinear legs, 
$k_{a_i} = z_i (k_{a_1}+\cdots+k_{a_m})$.
In the strongly-ordered limit, where $a_1$ is more nearly collinear with $a_2$
than either with $a_3$, and so on (that is, where 
$s_{a_1 a_2} \ll s_{a_2 a_3}, s_{a_1 a_2 a_3}$), the splitting amplitude also
simplifies.  Here, unlike the soft case, there is a sum over intermediate 
helicities,
$$\eqalign{
 &\Split^{\rm tree}_{-\lambda}(a_1^{\lambda_{a_1}},a_2^{\lambda_{a_2}},\ldots,
                               a_m^{\lambda_{a_m}};\{z_i\})
  \inlimit^{s_{a_1 a_2} \ll s_{a_1 a_2 a_3} \ll \cdots \ll s_{a_1\cdots a_m}}\cr
&\hskip 20mm
\sum_{\lambda_{12},\lambda_{123},\ldots = \pm}
 \Split^{\rm tree}_{-\lambda_{12}}(a_1^{\lambda_{a_1}},a_2^{\lambda_{a_2}};{z_1\over z_1+z_2})
 \Split^{\rm tree}_{-\lambda_{123}}((a_1+a_2)^{\lambda_{12}},a_3^{\lambda_{a_3}};
                                    {z_1+z_2\over z_1+z_2+z_3})
\cdots
\cr &\hskip 20mm\hphantom{ \sum_{\lambda_{12},\lambda_{123},\ldots = \pm} C}
\times \Split^{\rm tree}_{-\lambda_{12\cdots (m-1)}}(
            (a_1+\cdots+a_{m-1})^{\lambda_{12\cdots (m-1)}},a_m^{\lambda_{a_m}};
                                    1-z_m) + {\rm subleading}.
}\anoneqn$$

As an example,
the triply-collinear splitting amplitude~[\use\DDFM,\use\MultipleAntenna],
 $\Ctree_{+}(1^-,2^+,3^-;z_1,z_2)$, simplifies to
$$
\Ctree_{+}(R^+,3^-;z_R)\Ctree_{-}(1^-,2^+;\textstyle{z_1\over z_1+z_2})
+\Ctree_{+}(R^-,3^-;z_R)\Ctree_{+}(1^-,2^+;\textstyle{z_1\over z_1+z_2})
+ \cdots
\anoneqn$$
in the strongly-ordered limit.

Because of the summation over intermediate helicities, however, in general the
splitting function (the helicity-summed and -averaged square of the splitting
amplitude) will not factorize simply.  This can be seen, for example, by
examining the strongly-ordered limit of
the triply-collinear splitting function~[\use\GloverCampbell,\use\CataniGrazzini],
\def\spacer{\hphantom{\biggl[]}}
$$\eqalign{
2 &\biggl[ {(z_2 s_{123}-(1-z_3) s_{23})^2\over s_{12}^2 s_{123}^2 (1-z_3)^2}
 +{2 s_{23}\over s_{12} s_{123}^2} 
 +{3\over 2 s_{123}^2}
\cr &\spacer
 +{1\over s_{12} s_{123}} \Bigl( 
      {(1-z_3 (1-z_3))^2\over z_3 z_1 (1-z_1)} 
       -{2 (z_2^2+z_2 z_3+z_3^2)\over 1-z_3}
       +{(z_2 z_1-z_2^2 z_3-2)\over z_3 (1-z_3)}  \Bigr)
\cr &\spacer
 +{1\over 2 s_{12} s_{23}} 
    \Bigl( 3 z_2^2-{2 (2-z_1+z_1^2)(z_2^2+z_1 (1-z_1))\over z_3 (1-z_3)} 
           + {1\over z_1 z_3 }
           +{1\over (1-z_1)(1-z_3)} \Bigr) \biggr]
\cr &
+ (s_{12},z_1) \leftrightarrow (s_{23},z_3).
}\eqn\TriplyCollinear$$

We must be careful in taking this limit ($s_{12} \ll s_{23}, s_{123}$). Although 
$s_{23} \rightarrow z_2/(1-z_3) s_{123}$, so that the numerator of the first term
vanishes, the denominator contains a double pole in $s_{12}$, and so the relative
rate at which this limit is approached, compared to the vanishing of 
$s_{12}$, becomes important.

I will make use of the generalized Gram determinant $G$,
$$
G\biggl( {p_1,\ldots,p_n\atop q_1,\ldots,q_n}\biggr)
= \det( 2 p_i\cdot q_j ),
\eqn\GeneralizedGramDet$$
which vanishes whenever two $p_i$ or two $q_i$ become collinear (or
when any momentum becomes soft).
It will also be convenient to define
$$
\Delta(p_1,\ldots,p_n) \equiv 
G\biggl( {p_1,\ldots,p_n\atop p_1,\ldots,p_n}\biggr).
\eqn\GramDetDefinition$$
(The normalization in these definitions is non-standard.)

Using this generalized Gram determinant, we can rewrite the first term 
in~(\use\TriplyCollinear) as follows,
$$\eqalign{
{(z_2 s_{123}-(1-z_3) s_{23})^2\over s_{12}^2 s_{123}^2 (1-z_3)^2}
&= {(q\cdot k_2 s_{123} - q\cdot (k_1+k_2) s_{23})^2\over s_{12}^2 s_{123}^2 
       \LB q\cdot (k_1+k_2)\RB^2}
\cr &= {(G\L {1,2\atop q,3}\R - q\cdot k_2 s_{12})^2\over s_{12}^2 s_{123}^2 
       \LB q\cdot (k_1+k_2)\RB^2},
}\eqn\FirstTerm$$
where $q$ is a reference momentum --- a massless momentum not collinear to the $k_i$.

Using spinor products, we can see that
$$G\L {1,2\atop q,3}\R = s_{1q} s_{23}-s_{2q} s_{13}
= -\spa1.2\spa{3}.q\spb2.3\spb{1}.q
-\spb1.2\spb{3}.q\spa2.3\spa{1}.q
-s_{12} s_{3q}
\sim \sqrt{s_{12}}
\anoneqn$$
as $s_{12}\rightarrow 0$.

In the limit, eqn.~(\use\FirstTerm) thus becomes,
$$
{G^2\L {1,2\atop q,3}\R\over s_{12}^2 s_{123}^2 
       \LB q\cdot (k_1+k_2)\RB^2} + {\cal O}(s_{12}^{-1/2});
\eqn\FirstTermB$$
the subleading term leads to finite integrals over singular regions and is
thus not ultimately important to extracting poles in the dimensional regulator $\e$.

The limit of the remaining terms in eqn.~(\use\TriplyCollinear) is straightforward,
and we obtain
$$\eqalign{
&{G^2\L {1,2\atop q,3}\R\over s_{12}^2 s_{123}^2 
       \LB q\cdot (k_1+k_2)\RB^2} 
\cr & +{4 (1 - z_1 + z_1^2 - z_3 + z_1 z_3 + z_3^2) 
 \LB (1-z_3+z_3^2)^2 + (z_1^2+z_1 z_3-z_3) (1-z_3+z_3^2) -z_1 (1-z_1 z_3)\RB
\over s_{12} s_{123} z_1 (1 - z_3) z_3 (1 - z_1 - z_3)}
}\eqn\TriplyCollinearStronglyOrderedA$$ 
for the triply-collinear splitting function.  This is not equal to the
product of nested two-particle splitting functions,
$$
{4 (1-z_3+z_3^2)^2 (1 - z_1 + z_1^2 - 2 z_3 + z_1 z_3 + z_3^2)^2
\over s_{12} s_{123} z_1 (1-z_1-z_3) z_3 (1-z_3)^3}.
\anoneqn$$
We may observe, however, that the Gram determinant depends on the azimuthal angle
around the $k_1+k_2$ axis.  Nothing in any hard cross-section will depend on
this angle in the collinear limit.  We will eventually need to integrate over it
anyway; if we average over it, we can see that the Gram determinant 
averages to zero
(incidentally reducing the subleading term in eqn.~(\use\FirstTermB) to
${\cal O}(s_{12}^0)$), while its square averages to 
$$
2 s_{12} s_{23} s_{3q} s_{1q}
+s_{12}^2 s_{3q}^2.
\anoneqn$$
Plugging in this expression, eqn.~(\use\TriplyCollinearStronglyOrderedA) becomes
$$\eqalign{
&{ 4 (1-z_1-z_2) z_1 z_3 \over s_{12} s_{123} (1-z_3)^3}
+{4 (1 - z_1 + z_1^2 - z_3 + z_1 z_3 + z_3^2) 
\over s_{12} s_{123} z_1 (1 - z_3) z_3 (1 - z_1 - z_3)}
\cr &\hphantom{ { 4 (1-z_1-z_2) z_1 z_3 \over s_{12} s_{123} (1-z_3)^3} +}
\hskip 5mm\times  
 \LB (1-z_3+z_3^2)^2 + (z_1^2+z_1 z_3-z_3) (1-z_3+z_3^2) -z_1 (1-z_1 z_3)\RB
 + {\cal O}(s_{12}^0)
\cr &= 
{4 (1-z_3+z_3^2)^2 (1 - z_1 + z_1^2 - 2 z_3 + z_1 z_3 + z_3^2)^2
\over s_{12} s_{123} z_1 (1-z_1-z_3) z_3 (1-z_3)^3},
}\anoneqn$$
as desired.  Accordingly, if we redefine the averaging operation
$\langle\rangle$ to include not
only helicity summation and averaging but also averaging over azimuthal angles
of collinear pairs,
then we obtain 
$$\eqalign{
 &\bigl\langle|\Split^{\rm tree}_{-\lambda}(a_1^{\lambda_{a_1}},a_2^{\lambda_{a_2}},\ldots,
                               a_m^{\lambda_{a_m}};\{z_i\})|^2\bigr\rangle
  \inlimit^{s_{a_1 a_2} \ll s_{a_1 a_2 a_3} \ll \cdots \ll s_{a_1\cdots a_m}}\cr
&\hskip 20mm
 \bigl\langle|\Split^{\rm tree}_{-\lambda_{12}}(a_1^{\lambda_{a_1}},a_2^{\lambda_{a_2}};{z_1\over z_1+z_2})|^2\bigr\rangle
 \bigl\langle|\Split^{\rm tree}_{-\lambda_{123}}((a_1+a_2)^{\lambda_{12}},a_3^{\lambda_{a_3}};
                                    {z_1+z_2\over z_1+z_2+z_3})|^2\bigr\rangle
\cdots
\cr &\hskip 20mm\hphantom{ C}
\times \bigl\langle|\Split^{\rm tree}_{-\lambda_{12\cdots (m-1)}}(
            (a_1+\cdots+a_{m-1})^{\lambda_{12\cdots (m-1)}},a_m^{\lambda_{a_m}};
                                    1-z_m)|^2\bigr\rangle + {\rm subleading}
}\anoneqn$$
for the squared splitting function in the strongly ordered limit.

\section{Antenna Factorization}
\tagsection\AntennaReviewSection
\vskip 10pt

We can unify the soft and collinear limits by associating a single 
function~[\use\SingleAntenna]
with each color-connected triplet of momenta.  In the singular limit, the
triplet reduces to a pair of massless momenta.  We can remap the three momenta
to two massless momenta even away from the singular limit using a pair
of {\it reconstruction\/} functions.  For a triplet of momenta $(k_a,k_1,k_b)$,
with $k_{a,b}$ remaining hard in any singular limit under consideration,
the remappings to a massless pair $k_{\ah,\bh}$ are,
$$\eqalign{
k_\ah &= f_\ah(a,1,b) \equiv -{1\over 2 (K^2-s_{1b})} \LB {(1+\rho) K^2}
            - 2 s_{1b} r_1\RB\, k_a
          - r_1 k_1 
\cr &\hphantom{= f_\ah(a,1,b) = !}
    -{1\over 2 (K^2 - s_{a1})}\LB (1-\rho) K^2 -2 s_{1a} r_1\RB\, k_b\,,\cr
k_\bh &= f_\bh(a,1,b) \equiv
  -{1\over 2 (K^2 - s_{1b})} \LB (1-\rho)K^2 -2 s_{1b} (1-r_1)\RB\, k_a
          - (1-r_1) k_1 
\cr &\hphantom{= f_\ah(a,1,b) = !}
    -{1\over 2 (K^2 - s_{a1})}\LB (1+\rho)K^2 - 2 s_{1a} (1-r_1)\RB\, k_b\,,\cr
}\eqn\ReconstructionFunctions$$
where $K=k_a+k_1+k_b$, $r_1 = s_{1b}/(s_{a1}+s_{1b})$, and 
$$
\rho = \sqrt{1+ {4 r_1 (1-r_1) s_{1a} s_{1b}\over K^2 s_{ab}}}.
\anoneqn$$
(Other choices for $r_1$ are possible, subject to certain 
constraints~[\use\MultipleAntenna].)

\midinsert
\LoadFigure\AntennaDefinitionFigure
{\baselineskip 13 pt
\noindent\narrower
The antenna amplitude expressed in terms of Berends--Giele currents.
}  {\epsfysize 1.5 truein}{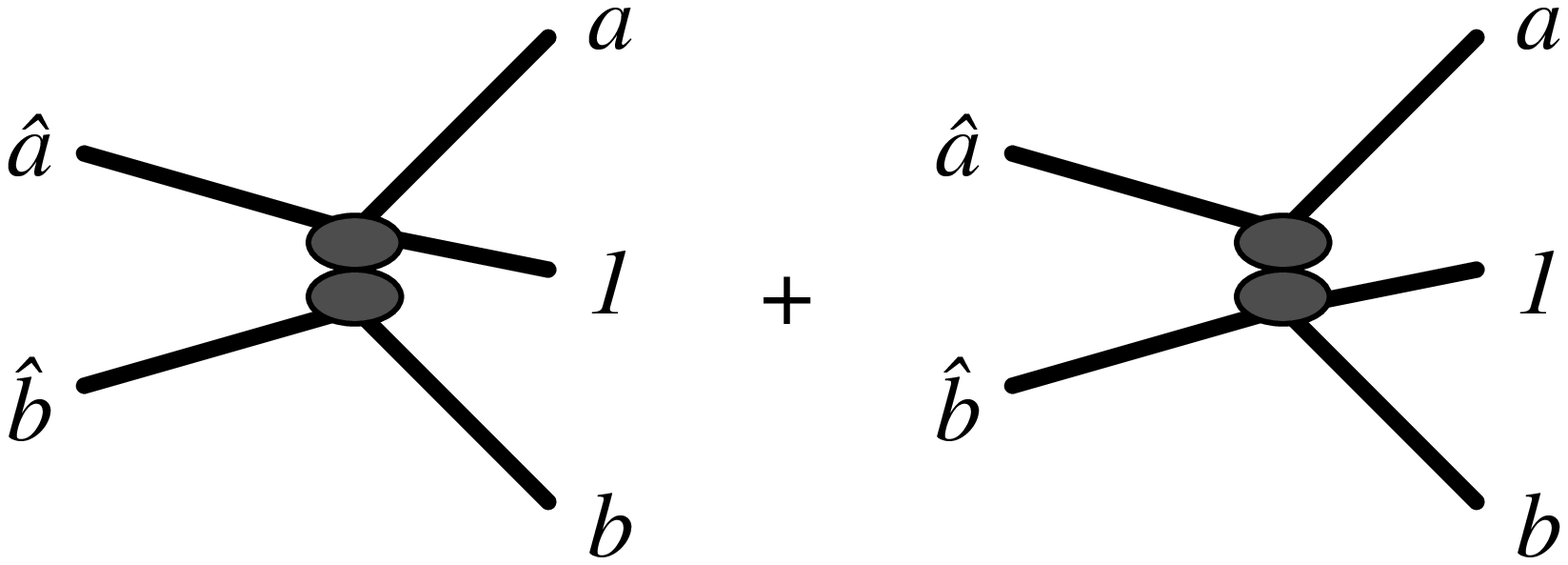}{}\endinsert

These reconstruction functions reduce to the usual combinations in the various
soft and collinear limits,
$$\eqalign{
k_\ah &= -(k_a\tp k_1), k_\bh = -k_b,\qquad {\rm when\ } k_a\parallel k_1, {\rm\ i.e.\ }
  s_{a1} = 0, s_{1b}\neq 0;
\cr
k_\ah &= -k_a, k_\bh = -(k_1\tp k_b),\qquad {\rm when\ } k_1\parallel k_b,
{\rm\ i.e.\ } s_{a1} \neq 0, s_{1b}= 0;
\cr
k_\ah &= -k_a, k_\bh = -k_b,\hphantom{({} \tp k_b)}\qquad
   {\rm when\ } k_1 {\rm\ is\ soft, i.e.\ } 
 s_{a1} = 0 =s_{1b}.
\cr}\anoneqn$$
\topinsert
\LoadFigure\AntennaFactorizationFigure
{\baselineskip 13 pt
\noindent\narrower
The factorization of a tree amplitude into an antenna amplitude and a hard amplitude.
The shaded circles represent sums over all tree diagrams.
}  {\epsfxsize 5.5 truein}{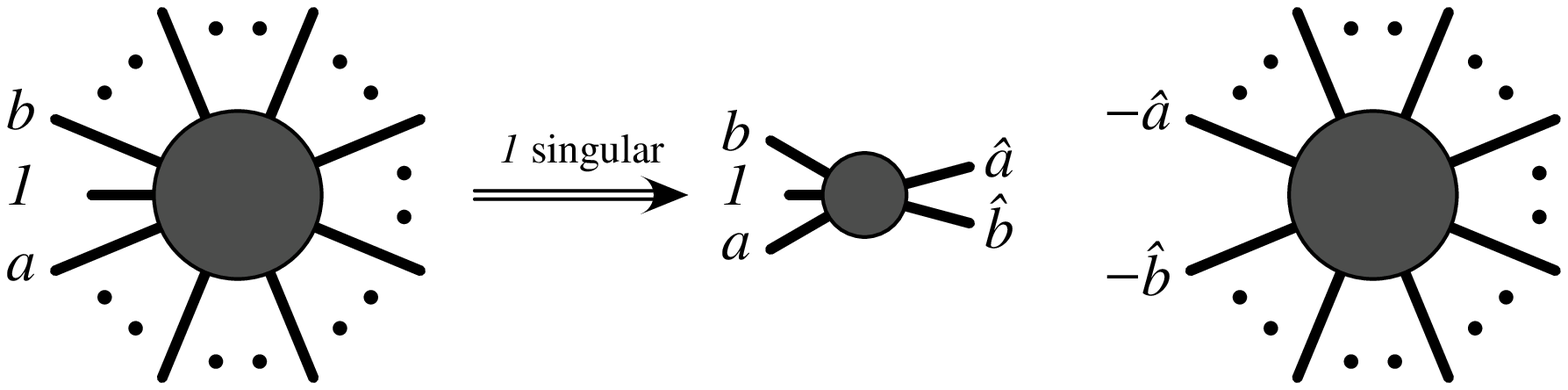}{}\endinsert
Beyond producing a pair of massless momenta satisfying momentum conservation,
\hbox{$K=-k_\ah-k_\bh$}, these reconstruction functions also ensure that the soft
and collinear limits can be unified into a single factorizing amplitude,
$$\eqalign{
&A_n(\ldots,a^{\lambda_a},1^{\lambda_1},b^{\lambda_b},\ldots) 
\inlimit^{k_1 {\rm\ singular}} \sum_{\phpol \lambda_{\ah,\bh}}\cr
&\hskip 10mm\Ant(\ah^{\lambda_\ah},\bh^{\lambda_\bh}\leftarrow 
 a^{\lambda_a},1^{\lambda_1},b^{\lambda_b})
A_{n-1}(\ldots,-k_{\ah}^{-\lambda_\ah},-k_{\bh}^{-\lambda_\bh},\ldots).
}\eqn\SingleEmissionFactorization$$
This factorization is depicted in \fig\AntennaFactorizationFigure.
The {\it antenna\/} amplitude has an explicit expression in terms of the
Berends--Giele current $J$~[\use\Recurrence,\use\DixonTASI,\use\HelicityRecurrence], 
$$
\Ant(\ah,\bh\leftarrow a,1,b) = 
J(a,1;\ah) J(b;\bh)+J(a;\ah) J(1,b;\bh).
\eqn\AntennaDef$$
The complete
list of its helicity amplitudes was given in ref.~[\use\MultipleAntenna].  It
is depicted diagrammatically in fig.~\AntennaDefinitionFigure.

We can use the generalized Gram determinant~(\use\GramDetDefinition)
to define the singular limit,
$$
k_1 {\rm\ singular\ } \Longleftrightarrow 
L(a,1,b) \equiv 
{1\over s_{ab}^3} G\biggl({a,1,b\atop a,1,b}\biggr) \rightarrow 0.
\eqn\SingleCriterion$$
I will denote this singular limit by $\sing{k_1}\rightarrow 0$.  This
gives a Lorentz-invariant definition of `soft' and `collinear'; in the
limit $k_a\parallel k_1$, $k_b$ is effectively the reference momentum defining
the transverse direction, and $k_a$ plays that role in the $k_b\parallel k_1$
limit.

\topinsert
\LoadFigure\DoubleAntennaDefinitionFigure
{\baselineskip 13 pt
\noindent\narrower
The double-emission antenna amplitude expressed in terms of Berends--Giele currents.
}  {\epsfysize 1.5 truein}{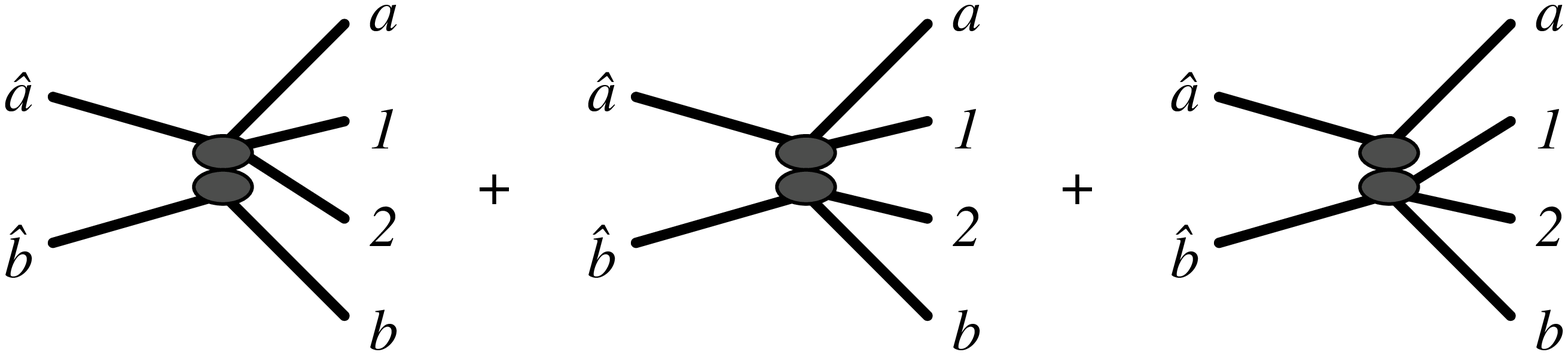}{}\endinsert
The reconstruction functions generalize to the emission of $n$ singular
particles, mapping $n+2$ momenta to two massless momenta.  (Explicit forms
of these functions suitable for uniform limits 
are given in ref.~[\use\MultipleAntenna]; for strongly-ordered limits, 
these must be generalized as discussed in the appendix.)  The definition
of the antenna function also generalizes,
$$\eqalign{
\Ant&(\ah,\bh\leftarrow a,1,\ldots,m,b) =
\sum_{j=0}^m J(a,1,\ldots,j;\ah^{\{j\}})
             J(j\tp1,\ldots,m,b;\bh^{\{j\}}),
\cr
}\eqn\MultipleEmissionAntennaDef$$
where in order to ensure appropriate strongly-ordered limits 
in the appendix, different interpolation functions $r_i^{\{j\}}$
leading to different $k_{\ah}$ and $k_{\bh}$ 
are chosen in different
terms.  The antenna amplitude on the left-hand side is still expressed in
terms of the original $\ah$ and $\bh$.
The corresponding factorization is,
$$\eqalign{
A_n&(\ldots,a,1,\ldots,m,b,\ldots) 
\inlimit^{\sing{k_1},\ldots,\sing{k_m} \rightarrow 0} \cr
&\hskip 10mm 
\sum_{\phpol \lambda_{\ah,\bh}}
\Ant(\ah^{\lambda_\ah},\bh^{\lambda_\bh}\leftarrow a,1,\ldots,m,b)
A_{n-m}(\ldots,-k_{\ah}^{-\lambda_\ah},-k_{\bh}^{-\lambda_\bh},\ldots).
}\eqn\MultipleEmissionAntennaFactorization$$
I have left the dependence on the helicities of the singular legs implicit.
The structure of the antenna amplitude for $m=2$ is depicted in 
fig.~\use\DoubleAntennaDefinitionFigure.

\section{Iterated Antenn\ae}
\tagsection\IteratedAntennaSection
\vskip 10pt

\topinsert\LoadFigure\NestedAntennaFigure
{\baselineskip 13 pt
\noindent\narrower
Factorization of the double-emission antenna amplitude in a strongly-ordered limit.
}  {\epsfysize 1.5 truein}{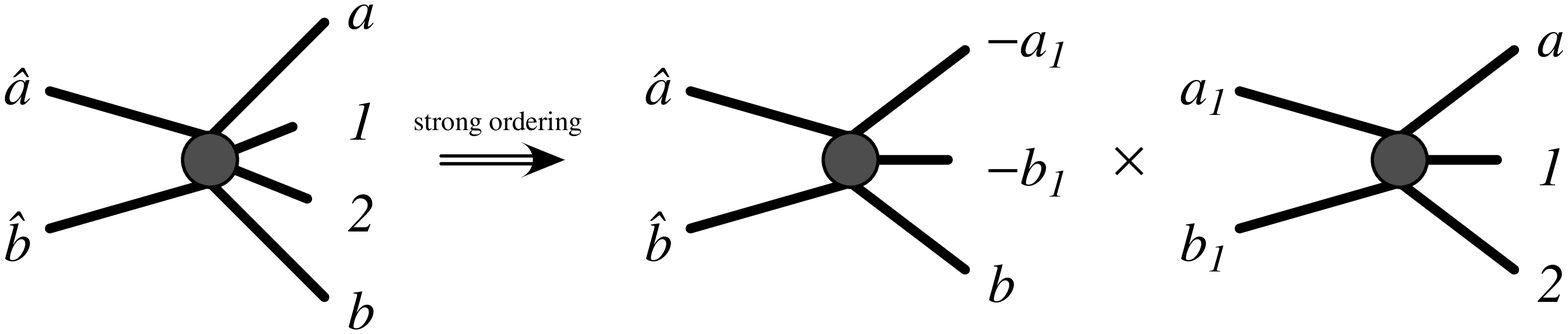}{}\endinsert

We may expect a simplification of the multiple-emission antenna 
function~(\use\MultipleEmissionAntennaDef) in the strongly-ordered limit,
where the singular momenta are strongly ordered in energy or angle.  Let us
postpone for a bit the question of what precisely we mean by `strong ordering'
when the two types of limits are intermixed.  Consider the emission of
two singular gluons.  
If the first, with momentum $k_1$,
 is more singular than the second, with momentum $k_2$ (that is, softer 
or more collinear to one of the hard legs), then at the first step, we can treat $k_2$
as another hard leg for the purposes of factorization.  This gives the following
factorization,
\def\ab#1{{a_{#1}}}
\def\bb#1{{b_{#1}}}
$$\eqalign{
A_n&(\ldots,a,1,2,b,\ldots) 
\inlimit^{}\cr
&\hskip 10mm \sum_{\phpol \lambda_{\ab1,\bb1}}
\Ant(\ab1^{\lambda_\ab1},\bb1^{\lambda_\bb1}\leftarrow a,1,2)
A_{n-1}(\ldots,-k_{\ab1}^{-\lambda_\ab1},-k_{\bb1}^{-\lambda_\bb1},b,\ldots).
}\anoneqn$$
Next we factorize the surviving amplitude, given that $(k_\ab1,k_\bb1,k_b)$ is also 
a singular configuration,
$$\eqalign{
A_n&(\ldots,a,1,2,b,\ldots) 
\inlimit^{}\cr
&\hskip 10mm \sum_{\phpol \lambda_{\ab1,\bb1,\ah,\bh}}
\Ant(\ab1^{\lambda_\ab1},\bb1^{\lambda_\bb1}\leftarrow a,1,2)
\Ant(\ah^{\lambda_\ah},\bh^{\lambda_\bh}\leftarrow 
     -k_\ab1^{-\lambda_\ab1},-k_\bb1^{-\lambda_\bb1},b)
\cr &\hphantom{ \sum_{\phpol \lambda_{\ab1,\bb1,\ah,\bh}} }\hskip 20mm\times
A_{n-2}(\ldots,-k_{\ah}^{-\lambda_\ah},-k_{\bh}^{-\lambda_\bh},\ldots).
}\anoneqn$$
Comparing with the double-singular antenna factorization as given by 
eqn.~(\use\MultipleEmissionAntennaFactorization),
$$\eqalign{
A_n&(\ldots,a,1,2,b,\ldots) 
\inlimit^{\sing{k_1},\sing{k_2} \rightarrow 0} \cr
&\hskip 10mm 
\sum_{\phpol \lambda_{\ah,\bh}}
\Ant(\ah^{\lambda_\ah},\bh^{\lambda_\bh}\leftarrow a,1,2,b)
A_{n-2}(\ldots,-k_{\ah}^{-\lambda_\ah},-k_{\bh}^{-\lambda_\bh},\ldots),
}\anoneqn$$
we see that in the strongly-ordered limit,
$$\eqalign{
\Ant&(\ah^{\lambda_\ah},\bh^{\lambda_\bh}\leftarrow a,1,2,b) 
\inlimit^{\sing{k_1} \ll \sing{k_2} \rightarrow 0}\cr
&\hskip 10mm \sum_{\phpol \lambda_{\ab1,\bb1}}
\Ant(\ab1^{\lambda_\ab1},\bb1^{\lambda_\bb1}\leftarrow a,1,2)
\Ant(\ah^{\lambda_\ah},\bh^{\lambda_\bh}\leftarrow 
     -k_\ab1^{-\lambda_\ab1},-k_\bb1^{-\lambda_\bb1},b)
+\cdots
}\eqn\StronglyOrdered$$
This factorization is shown schematically in fig.~\use\NestedAntennaFigure.
In these equations, the momenta $\ab1$ and $\bb1$ are reconstructed using
the single-emission functions~(\use\ReconstructionFunctions),
$$\eqalign{
k_\ab1 &= f_\ah(a,1,2),\cr
k_\bb1 &= f_\bh(a,1,2).\cr
}\eqn\InnerMomenta$$
The final momenta $\ah$ and $\bh$ can be defined either by iterating the
single-emission reconstruction functions (as they naturally would be on the
right-hand side),
$$\eqalign{
k_\ah &= f_\ah(-k_{\ab1},-k_{\bb1},b),\cr
k_\bh &= f_\bh(-k_{\ab1},-k_{\bb1},b),\cr
}\anoneqn$$
or directly using the multiple-emission reconstruction functions of
ref.~[\use\MultipleAntenna],
$$\eqalign{
k_\ah &= f_\ah(a,1,2,b),\cr
k_\bh &= f_\bh(a,1,2,b),\cr
}\anoneqn$$
as they naturally would be on the left-hand side of eqn.~(\use\StronglyOrdered).
While the two sets of definitions are not identical, the differences will give
rise only to subleading terms.

The notation $\sing{k_1}\ll \sing{k_2}$ means that $k_1$ is (much) more
singular than $k_2$, that is that they are strongly ordered; 
but what precisely do we mean by this statement?  One might try to use the 
single-emission criterion~(\use\SingleCriterion) not only to define
soft or collinear momenta, but also to compare the relative degree of
softness or collinearity.  However, if we use the nested antenn\ae{} as
arguments, the strong-ordering criterion would be,
$$
L(a,1,2) \ll L(\ab1,\bb1,b)
\anoneqn$$
which would require 
$$
E_1 \ll E_2 \sqrt{{E_2\over E_b}},
\anoneqn$$
too stringent a requirement for the strongly-ordered soft limit.
In contrast, if we require 
$$
L(a,1,b) \ll L(a,2,b),
\anoneqn$$
then the soft limits are distinguished properly, but the collinear
limit in which $\theta_{12} \ll \theta_{a1},\theta_{a2}$, where the
strongly-ordered form~(\use\StronglyOrdered) also holds, is not
picked up properly.  The complete set of limits in which the 
simplification~(\use\StronglyOrdered) holds is,
$$\eqalign{
E_1 &\ll E_2 \quad(k_{1,2} {\rm\ soft});\cr
\theta_{a1} &\ll \theta_{a2}, \theta_{12} \quad(a\parallel 1\parallel 2);\cr
\theta_{12} &\ll \theta_{a1}, \theta_{a2} \quad(a\parallel 1\parallel 2);\cr
\theta_{a1} &\ll \theta_{2b} \quad(a\parallel 1, 2\parallel b);\cr
}\anoneqn$$
and $k_1$ soft with $k_2\parallel k_b$.

We need a more complicated function to discriminate between these limits
and other singular regions; the following form will work,
$$ 
L_s(a,1,2,b) \equiv
\LB {\displaystyle G\biggl({a,1,b\atop a,1,b}\biggr) 
      G\biggl({a,2,b\atop a,2,b}\biggr) - G^2\biggl({a,1,b\atop a,2,b}\biggr)
  \over\displaystyle G^2\biggl({a, k_1+k_2,b\atop a, k_1+k_2,b}\biggr)}
\RB^p
\eqn\LsDef$$
(it is of course not unique), where $p>0$ will be chosen as described in
the following.  

We may note that in double-soft limits, the numerator scales as $E_1^2 E_2^2$,
while the denominator scales as $(E_1+E_2)^4$; so in either strongly-ordered
limit $E_1\ll E_2$ or $E_2\ll E_1$, $L_s$ will tend to zero.  
In the limit $\theta_{a1}\ll \theta_{a2},\theta_{12}\sim\theta \ll 1$, 
the denominator
scales like $\theta^4$, while the numerator scales like 
$\theta_{a1}^2\theta^2$, so again it tends to zero.  Finally, in
the limit $\theta_{12}\ll \theta_{a1}, \theta_{a2}\sim \theta \ll 1$,
the denominator again scales like $\theta^4$, while the numerator
scales like $\theta_{12}^2\theta^2$, so once again $L_s$ tends to zero.

However, in unordered double-soft limits $E_1\sim E_2\ll E_{a,b}$, this
function is also rather small (though tending to a constant rather than to
zero).  In order to better distinguish this region from
the strongly-ordered collinear limits, we may choose $p<1$; empirically,
$p=1/4$ is a good choice.

The discriminant $L_s$ is symmetric in $k_1\leftrightarrow k_2$; accordingly,
it distinguishes regions where either $k_1$ or $k_2$ is much more singular
than the other from regions where both are comparably singular.  It does not,
however, distinguish the two different limits ($\sing{k_1}\ll \sing{k_2}$ and
$\sing{k_2}\ll \sing{k_1}$) from each other.  To do that, introduce another
function,
$$
L_r(a,b;1;2) = {s_{a1} s_{12b}\over s_{a12} s_{2b}},
\eqn\LrDef$$
where $s_{ijl} = (k_i+k_j+k_l)^2$.
This function tends to zero when $(a,1,2)$ should be taken to be the inner
(i.e. more singular)
antenna, and to infinity when $(1,2,b)$ should be taken to be the inner antenna.
For configurations where $k_1\parallel k_a$ and $k_2\parallel k_b$
with $\theta_{a1}\sim \theta_{2b}$, it is
of order unity, and either antenna may be taken to be the inner one.  In this
latter case, we can establish an arbitrary boundary between the two factorizations
at $L_r = 1$.  We may summarize the constraints via the following equivalences,
$$\eqalign{
\sing{k_j} &\ll 1 \Longleftrightarrow
L(a,j,b) \ll 1;\cr
\sing{k_1} &\ll \sing{k_2}\Longleftrightarrow
L_s(a,1,2,b), L_r(a,b;1;2) \ll 1.
}\anoneqn$$

In considering the amplitude squared, we must again generalize the averaging
operation $\langle\rangle$, to include averaging over all azimuthal angles in
the different collinear limits.  In particular, this means averaging over variables
which interpolate between azimuthal angles around $k_a$ (for the
triply-collinear limit $k_a\parallel k_1\parallel k_2$) to angles around $k_b$
(for the
triply-collinear limit $k_1\parallel k_2\parallel k_b$).  To construct such
a variable, we need to introduce an additional, reference, momentum.  Roughly
speaking, we are already using $k_b$ as a reference momentum to
define the $a\parallel 1$ limit, so we need an additional momentum to define
an azimuthal angle.  Choose this additional momentum $q$ to be massless; we
can then define a variable,
$$\eqalign{
u &= {G\biggl( {q,a,b\atop a,1,b}\biggr)\over \sqrt{\Delta(q,a,1)\Delta(q,b,1)}}
\cr &= {s_{a1} s_{bq} + s_{1b} s_{aq} - s_{1q} s_{ab}\over
    2\sqrt{s_{a1} s_{1b} s_{aq} s_{bq}}}.
\cr
}\anoneqn$$
In the center-of-mass frame of $\ah$ and $\bh$, it corresponds to the cosine
of the azimuthal angle of $k_1$~[\use\Byckling].  Alternatively, with
$\varepsilon(1,2,3,4) = \epsilon_{\mu\nu\lambda\rho} 
k_1^\mu k_2^\nu k_3^\lambda k_4^\rho$,
we could use
$$\eqalign{
u &= 4\sqrt{-\Delta(a,b)\over\Delta(a,b,1)\Delta(a,b,q)}\varepsilon(q,a,1,b)
\cr &= {2\varepsilon(q,a,1,b)\over 
    \sqrt{s_{a1} s_{1b} s_{aq} s_{bq}}},
}\anoneqn$$
which is equal to the sine of the same angle.  
A third alternative is the phase,
$$
u = {\spa{a}.1\spa1.b\over \sqrt{s_{a1} s_{1b}}}
\anoneqn$$
where the reference momentum $q$ is implicit in the definition of the spinor
product.

Once we perform the angular averaging by integrating over this
variable,
the strongly-ordered factorization~(\use\StronglyOrdered) carries over to the
squared antenna functions,
$$\eqalign{
&\left\langle\LV\Ant(\ah,\bh\leftarrow a,1,2,b)\RV^2\right\rangle
\inlimit^{\sing{k_1} \ll \sing{k_2} \rightarrow 0}\cr
&\hskip 15mm 
\left\langle\LV\Ant(\ab1,\bb1\leftarrow a,1,2)\RV^2\right\rangle
\left\langle\LV\Ant(\ah,\bh\leftarrow 
     -k_\ab1,-k_\bb1,b)\RV^2\right\rangle + {\rm\ subleading}
}\eqn\StronglyOrderedSquare$$

The simplifications in the strongly-ordered limit generalize to the emission
of $m$ singular momenta, both at the amplitude level,
$$\eqalign{
\Ant&(\ah^{\lambda_\ah},\bh^{\lambda_\bh}\leftarrow a,1,\ldots,m,b) 
\inlimit^{\sing{k_1} \ll \sing{k_2}\ll\cdots\ll\sing{k_m} \rightarrow 0}\cr
&\hskip 10mm \sum_{\phpol \lambda_{\ab1,\bb1,\ldots,\ab{m},\bb{m}}}
\Ant(\ab1^{\lambda_\ab1},\bb1^{\lambda_\bb1}\leftarrow a,1,2)
\cr &\hskip 10mm\hphantom{ \sum_{\phpol \lambda_{\ab1,\bb1,\ldots,\ab{m},\bb{m}}} }
\hskip 5mm\times
\Ant(\ab2^{\lambda_\ab2},\bb2^{\lambda_\bb2}\leftarrow 
     -k_\ab1^{-\lambda_\ab1},-k_\bb1^{-\lambda_\bb1},3)
\cr &\hskip 10mm\hphantom{ \sum_{\phpol \lambda_{\ab1,\bb1,\ldots,\ab{m},\bb{m}}} }
\hskip 5mm\times\cdots\times
 \Ant(\ah^{\lambda_\ah},\bh^{\lambda_\bh}\leftarrow 
     -k_\ab{m-1}^{-\lambda_\ab{m-1}},-k_\bb{m-1}^{-\lambda_\bb{m-1}},b)
+\cdots
}\eqn\MStronglyOrdered$$
and at the level of squared amplitudes,
$$\eqalign{
&\left\langle\LV\Ant(\ah,\bh\leftarrow a,1,\ldots,m,b)\RV^2\right\rangle
\inlimit^{\sing{k_1} \ll \sing{k_2}\ll\cdots\ll\sing{k_m} \rightarrow 0}\cr
&\hskip 15mm 
\left\langle\LV\Ant(\ab1,\bb1\leftarrow a,1,2)\RV^2\right\rangle
\left\langle\LV\Ant(\ab2,\bb2\leftarrow -k_\ab1,-k_\bb1,3)\RV^2\right\rangle 
\cr &\hskip 25mm\times
\cdots\times
\left\langle\LV\Ant(\ah,\bh\leftarrow -k_\ab{m-1},-k_\bb{m-1},b)\RV^2\right\rangle 
+ {\rm\ subleading}
}\eqn\MStronglyOrderedSquare$$
In these equations, the momenta $\ab{j}$ and $\bb{j}$ are defined in nested form,
with $\ab1$ and $\bb1$ given by eqn.~(\use\InnerMomenta), and
$$\eqalign{
k_\ab{j} &= f_\ah(-k_{\ab{j-1}},-k_{\bb{j-1}},b),\cr
k_\bb{j} &= f_\bh(-k_{\ab{j-1}},-k_{\bb{j-1}},b).\cr
}\eqn\NestedMomenta$$

\section{Factorization in Loop Amplitudes}
\tagsection\LoopAntennaSection
\vskip 10pt

The soft and collinear factorization of tree amplitudes described in
sections~\use\MultipleSoftSection{} and~\use\MultipleCollinearSection{}
can be extended to loop corrections as well.  
\def\Gr{{\rm Gr}}
At one loop, the color decomposition analogous to~(\use\TreeColorDecomposition) is
$$
{\cal A}_n^\oneloop\L \{k_i,\lambda_i,a_i\}\R =
g^n \sum_J n_J
  \sum_{c=1}^{\lfloor{n/2}\rfloor+1}
      \sum_{\sigma \in S_n/S_{n;c}}
     \Gr_{n;c}\L \sigma \R\,A_{n;c}^{[J]}(\sigma),
\eqn\LoopColorDecomposition$$
where ${\lfloor{x}\rfloor}$ is the largest integer less than or equal to $x$
and $n_J$ is the number of particles of spin $J$.
The leading color-structure factor,
$$
\Gr_{n;1}(1) = N_c\ \Tr\L T^{a_1}\cdots T^{a_n}\R\,,
\anoneqn
$$
is just $N_c$ times the tree color factor, and the subleading color
structures are given by
$$
\Gr_{n;c}(1) = \Tr\L T^{a_1}\cdots T^{a_{c-1}}\R\,
\Tr\L T^{a_c}\cdots T^{a_n}\R.
\anoneqn
$$
$S_n$ is the set of all permutations of $n$ objects,
and $S_{n;c}$ is the subset leaving $\Gr_{n;c}$ invariant.
The decomposition~(\use\LoopColorDecomposition) holds separately
for different spins circulating around the loop.  The usual
normalization conventions take each massless spin-$J$ particle to have two
(helicity) states: gauge bosons, Weyl fermions, and complex scalars.
(For internal particles in the fundamental ($N_c+\bar{N_c}$) representation,
only the single-trace color structure ($c=1$) would be present,
and the corresponding color factor would be smaller by a factor of $N_c$.)

The subleading-color amplitudes $A_{n;c>1}$ are in fact not independent
of the leading-color amplitude $A_{n;1}$.  Rather, they can be expressed as
sums over permutations of the arguments of the latter~[\use\SusyFour].
(For amplitudes with external fermions, the basic objects are primitive
amplitudes~[\use\qqggg] rather than the leading-color one, but the same dependence
of the subleading color amplitudes holds.)

The leading-color amplitude at one loop factorizes as follows,
\def\spacer{\hphantom{ \inlimit^{k_s\rightarrow 0} !}}
$$\eqalign{
A_n^{\oneloop}(1,\ldots,a,s^\ls,b,\ldots,n)  &\inlimit^{k_s\rightarrow 0}
  \Soft^\tree(a,s^{\lambda_s},b)\,
      A_{n-1}^\oneloop(1,\ldots,a,b,\ldots,n)
\cr &\spacer
  +\Soft^\oneloop(a,s^{\lambda_s},b)\,
      A_{n-1}^\tree(1,\ldots,a,b,\ldots,n),
\cr  A_n^{\oneloop}(1,\ldots,a^\la,b^\lb,\ldots,n) 
 &\inlimit^{a \parallel b}\cr
 &\sum_{\phpol\ \lambda}  \biggl(
  \Split^\tree_{-\lambda}(a^{\la},b^{\lb})\,
      A_{n-1}^\oneloop(1,\ldots,(a+b)^\lambda,\ldots,n)
\cr &\hphantom{ \sum_{\phpol\ \lambda}  \biggl() }
  +\Split^\oneloop_{-\lambda}(a^\la,b^\lb)\,
      A_{n-1}^\tree(1,\ldots,(a+b)^\lambda,\ldots,n) \biggr).
}\eqn\OneLoopFactorization$$
in the soft and collinear 
limits~[\use\SusyFour,\use\qqggg,%
\use\BernChalmers,\use\AllOrdersCollinear], respectively.
(Explicit expressions for the one-loop factorizing amplitudes
$\Soft^\oneloop$ and $\Split^\oneloop$ were also computed to 
all orders in $\e$ in refs.~[\use\OneloopSplitB,\OneloopSplitA].)

At higher loops, the color decomposition similar
to eqns.~(\use\TreeColorDecomposition,\use\LoopColorDecomposition)
acquires more terms with additional traces, but the 
term leading in the number of colors has the form,
$$
{\cal A}_n^{l\lloop}\L \{k_i,\lambda_i,a_i\}\R =
g^{n+2l-2} N_c^l \sum_{\sigma \in S_n/Z_n}
\Tr(T^{a_{\sigma(1)}}\cdots T^{a_{\sigma(n)}})\,A_{n}^{{\rm LC}}(\sigma).
\eqn\LLoopColorDecomposition$$
(For fixed-order calculations, a different color 
decomposition~[\use\AlternateColorDecomposition] may be desirable.)

The antenna factorization of tree amplitudes described in 
section~\use\AntennaReviewSection{} can be extended~[\use\HigherLoopAntenna]
 to these leading-color
amplitudes, unifying the limits~(\use\OneLoopFactorization) given above.
To do so, first define a loop generalization of the Berends-Giele current.
Such a higher-loop current can be defined in terms of its 
unitarity cuts~[\use\SusyFour,\use\UnitarityB] to all orders 
in $\e$~[\use\BernMorgan,\use\UnitarityReview,\use\AllOrdersCollinear],
$$\eqalign{
&\LP J^{l\lloop}(1^\lu1,2^\lu2,\ldots,m^\lu{m};P)
   \RV_{t_{c\cdots d} {\rm\ cut}} =
\cr &\hskip 5mm\sum_{k=0}^{l-1} \,\sum_{j=2}^{l+ 1- k} 
\sum_{\phpol\ \sigma_i}\int d^{4-2\e}\LIPS(\ell_1,\ldots,\ell_j)\;
\cr &\hskip 5mm
  \hphantom{ \sum_{k=0}^{l-1}\,\sum_{j=2}^{l+ 1- k}\sum_{\phpol\ \sigma_i} }
\hskip 10mm \times J^{k\lloop}(1^\lu1,\ldots,(c\tm1)^\lu{c\tm1},
             \ell_1^{-\sigma_1},\ldots,\ell_j^{-\sigma_j},(d\tp1)^\lu{d\tp1},
             \ldots,m^{\lu{m}};P)
\cr &\hskip 75mm\times
   \,A_{d- c+ j+1}^{(l+ 1- j- k)\lloop}
  (c^\lu{c},\ldots,d^\lu{d},-\ell_j^{\sigma_j},\ldots,-\ell_1^{\sigma_1}).
}\eqn\HigherOrderJ$$
where $X^{0\lloop}$ means $X^{\tree}$, and $d^D\LIPS$,
the $D$-dimensional Lorentz-invariant phase-space measure.
(See ref.~[\use\CataniGrazziniSoft] for a related construction at one loop.)
While the currents appearing here must be evaluated in light-cone
gauge~[\use\MultipleAntenna,%
\use\CataniGrazzini,%
\ref\LanceComment{L. Dixon, personal communication}], 
the on-shell amplitudes on the other side of the cut may be
evaluated in any gauge.  This equation holds to all orders in $\e$;
for an $n$-loop antenna function for $m$ singular momenta, 
the $l$-loop current should be evaluated to ${\cal O}(\e^{2(n+m-l)})$,
with epsilonic powers of singular invariants left {\it un\/}expanded.
For example,
the cuts entering into the calculation of the one-loop four-point current are shown
in \fig\LoopCurrentDefinitionFigure.
\topinsert \LoadFigure\LoopCurrentDefinitionFigure
{\baselineskip 13 pt
\noindent\narrower
The different cuts required for the computation of the one-loop four-point current: 
(a) the cut in the $s_{123}$ channel (b) the cut in the $s_{12}$ channel.  The cut
in the $s_{23}$ channel is related to the latter by symmetry.  The quantities
on the left-hand side of the cuts
are currents, while those on the right-hand side
are on-shell amplitudes.
}{\epsfysize 1.5truein}{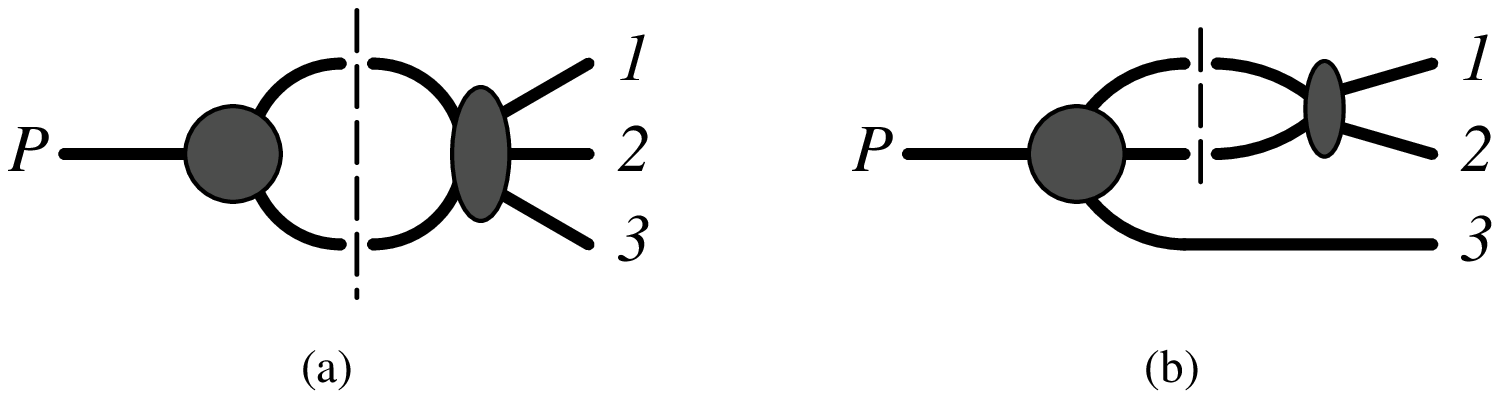}{}\endinsert

The use of cuts to compute amplitudes or currents can be
thought of as applying a modern version of dispersion relations.  There are
several significant differences from traditional versions of dispersion relations
that are worth keeping in mind.  Because of the use of dimensional regularization,
the integrals involve effectively {\it converge\/}, and so the subtraction ambiguities
that traditionally plagued dispersion relations are absent.  (It is for this reason
that the amplitudes must be kept to higher order in $\e$.)  It should also be
stressed that we do not want to apply the method to {\it integrals\/}
(although it is useful here too~[\use\vanNeerven]), but rather to 
{\it amplitudes\/}, so that we automatically take advantage 
of all the cancellations that 
have taken place inside lower-order quantities in a gauge-theory calculation.

\topinsert
\LoadFigure\LoopAntennaDefinitionFigure
{\baselineskip 13 pt
\noindent\narrower
Definition of the one-loop double-emission antenna amplitude in terms of currents.
The holes represent loops, the shaded circles sums over trees.
}  {\epsfysize 3.1 truein}{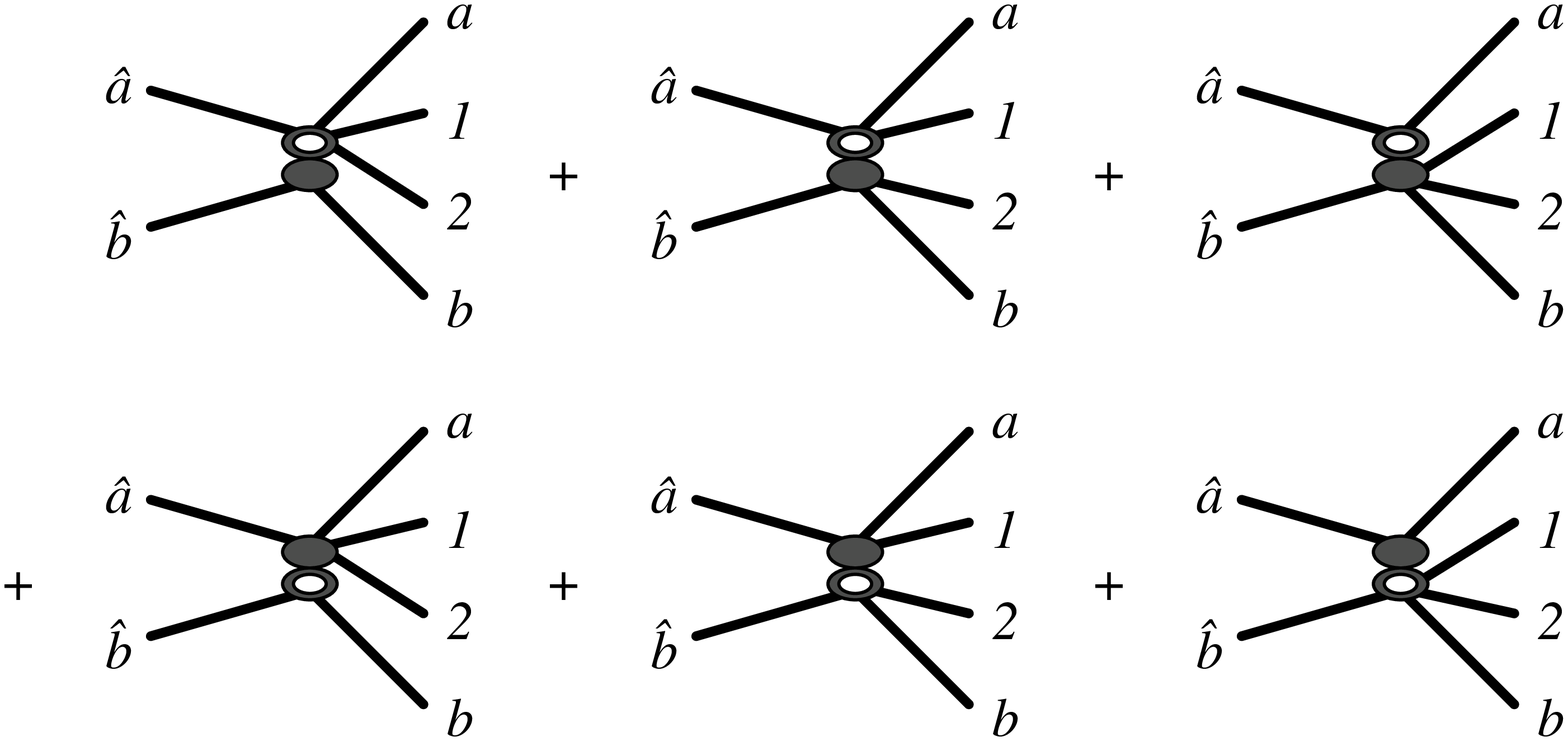}{}\endinsert

Using the higher-order current $J^{l\lloop}$, we can write down an
expression for the higher-loop generalization of the antenna amplitude,
$$
\Ant^{n\lloop}(\ah,\bh\leftarrow a,1,\ldots,m,b) =
\sum_{j=0}^m \sum_{l=0}^n J^{l\lloop}(a,1,\ldots,j;\ah^{\{j\}})
             J^{(n-l)\lloop}(j\tp1,\ldots,m,b;\bh^{\{j\}}).
\eqn\HigherLoopAntennaDef$$
The definition of the one-loop double-emission antenna
is shown as an example in \fig\LoopAntennaDefinitionFigure.
The factorization corresponding to eqn.~(\use\HigherLoopAntennaDef) is,
$$\eqalign{
&A_n^{r\lloop}(\ldots,a,1,\ldots,m,b,\ldots) 
\inlimit^{k_1,\ldots,k_m {\rm\ singular}}
\cr &\hskip 15mm
\sum_{\phpol\ \lambda_{\ah,\bh}} \sum_{v=0}^r 
  \Ant^{v\lloop}(\ah^{\lambda_\ah},\bh^{\lambda_\bh}\leftarrow a,1,\ldots,m,b) 
   A_{n-m}^{(r-v)\lloop}(\ldots,-k_\ah^{-\lambda_\ah},
                 -k_\bh^{-\lambda_\bh},\ldots).
}\eqn\MultipleEmissionHigherLoopAntennaFactorization$$
\topinsert
\LoadFigure\AntennaLoopFactorizationFigure
{\baselineskip 13 pt
\noindent\narrower
The factorization of a one-loop amplitude into antenna and hard amplitudes.
}  {\epsfxsize 5.5 truein}{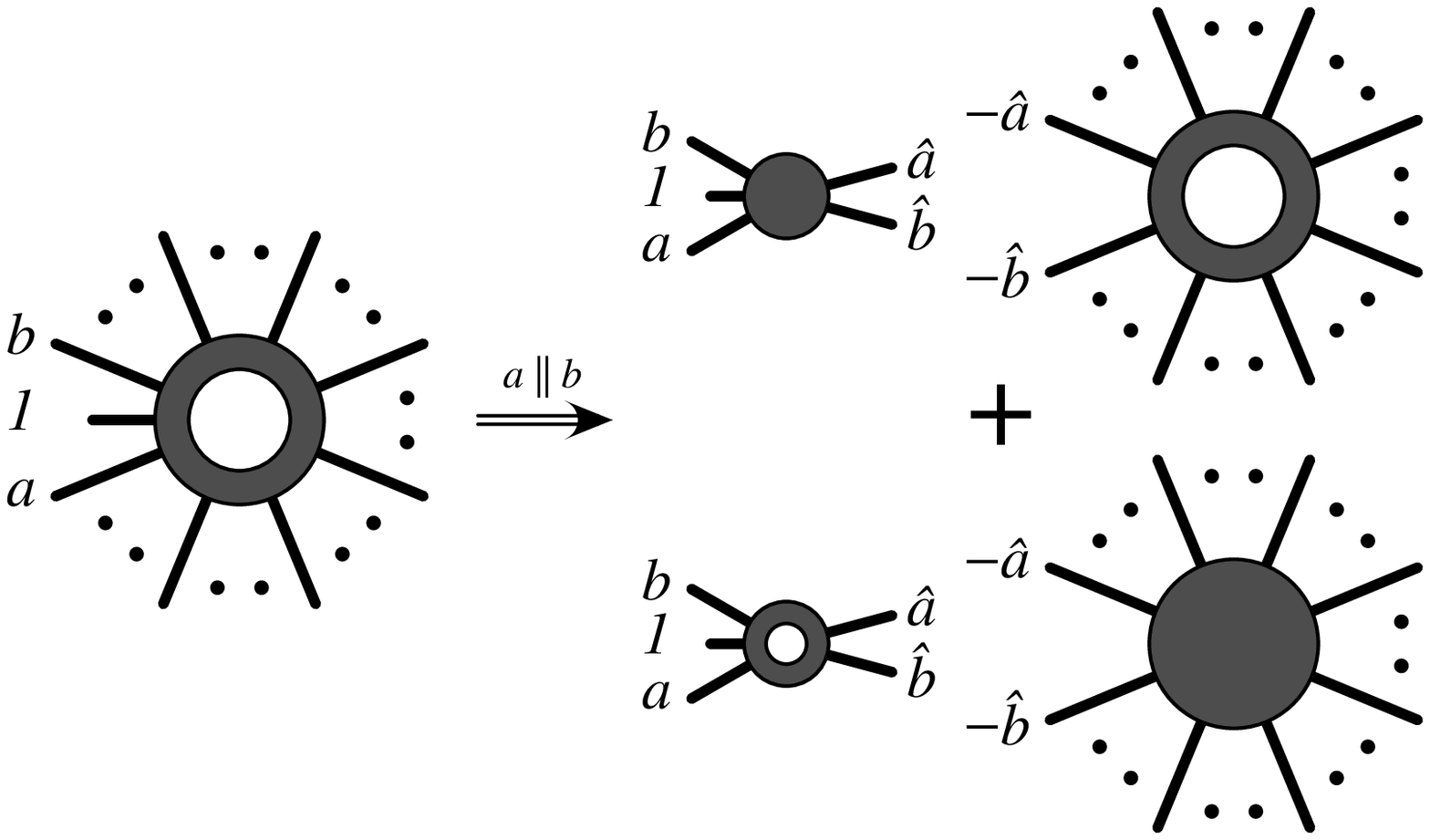}{}\endinsert
In the one-loop
case, this is a sum of two terms, similar to the factorizations
in eqn.~(\use\OneLoopFactorization); for example,
$$\eqalign{
&A_n^{\oneloop}(\ldots,a,1,b,\ldots) 
\inlimit^{\sing{k_1} \rightarrow 0}
\cr &\hskip 15mm
\sum_{\phpol\ \lambda_{\ah,\bh}} \Biggl(
  \Ant^{\tree}(\ah^{\lambda_\ah},\bh^{\lambda_\bh}\leftarrow a,1,b) 
   A_{n-m}^{\oneloop}(\ldots,-k_\ah^{-\lambda_\ah},
                 -k_\bh^{-\lambda_\bh},\ldots)
\cr &\hskip 15mm\hphantom{ \sum_{\phpol\ \lambda_{\ah,\bh}} \Biggl() }
  +\Ant^{\oneloop}(\ah^{\lambda_\ah},\bh^{\lambda_\bh}\leftarrow a,1,b) 
   A_{n-m}^{\tree}(\ldots,-k_\ah^{-\lambda_\ah},
                 -k_\bh^{-\lambda_\bh},\ldots)\Biggr),
}\eqn\SingleEmissionOneLoopAntennaFactorization$$
as depicted in \fig\AntennaLoopFactorizationFigure.

If we iterate this factorization, and 
compare with the direct factorization
of double-singular emission, and match coefficients of $A^\tree_{n-2}$, we
find that in the strongly-ordered limit,
$$\eqalign{
\Ant^\oneloop&(\ah^{\lambda_\ah},\bh^{\lambda_\bh}\leftarrow a,1,2,b) 
\inlimit^{\sing{k_1} \ll \sing{k_2} \rightarrow 0}\cr
&\hskip -15mm \sum_{\phpol \lambda_{\ab1,\bb1}} \biggl(
\Ant^\tree(\ab1^{\lambda_\ab1},\bb1^{\lambda_\bb1}\leftarrow a,1,2)
\Ant^\oneloop(\ah^{\lambda_\ah},\bh^{\lambda_\bh}\leftarrow 
     -k_\ab1^{-\lambda_\ab1},-k_\bb1^{-\lambda_\bb1},b)
\cr &\hskip -15mm \hphantom{ \sum_{\phpol \lambda_{\ab1,\bb1}} \Biggl() }
+\Ant^\oneloop(\ab1^{\lambda_\ab1},\bb1^{\lambda_\bb1}\leftarrow a,1,2)
\Ant^\tree(\ah^{\lambda_\ah},\bh^{\lambda_\bh}\leftarrow 
     -k_\ab1^{-\lambda_\ab1},-k_\bb1^{-\lambda_\bb1},b)
\biggr)
+\cdots
}\eqn\StronglyOrderedOneLoop$$
where the momenta are defined in eqn.~(\use\InnerMomenta).
(Matching coefficients of $A^\oneloop_{n-2}$ just reproduces 
eqn.~(\use\StronglyOrdered).)  This factorization is depicted in
\fig\NestedLoopAntennaFigure.

\topinsert\LoadFigure\NestedLoopAntennaFigure
{\baselineskip 13 pt
\noindent\narrower
Factorization of the one-loop
double-emission antenna amplitude in a strongly-ordered limit.
}  {\epsfysize 3.1 truein}{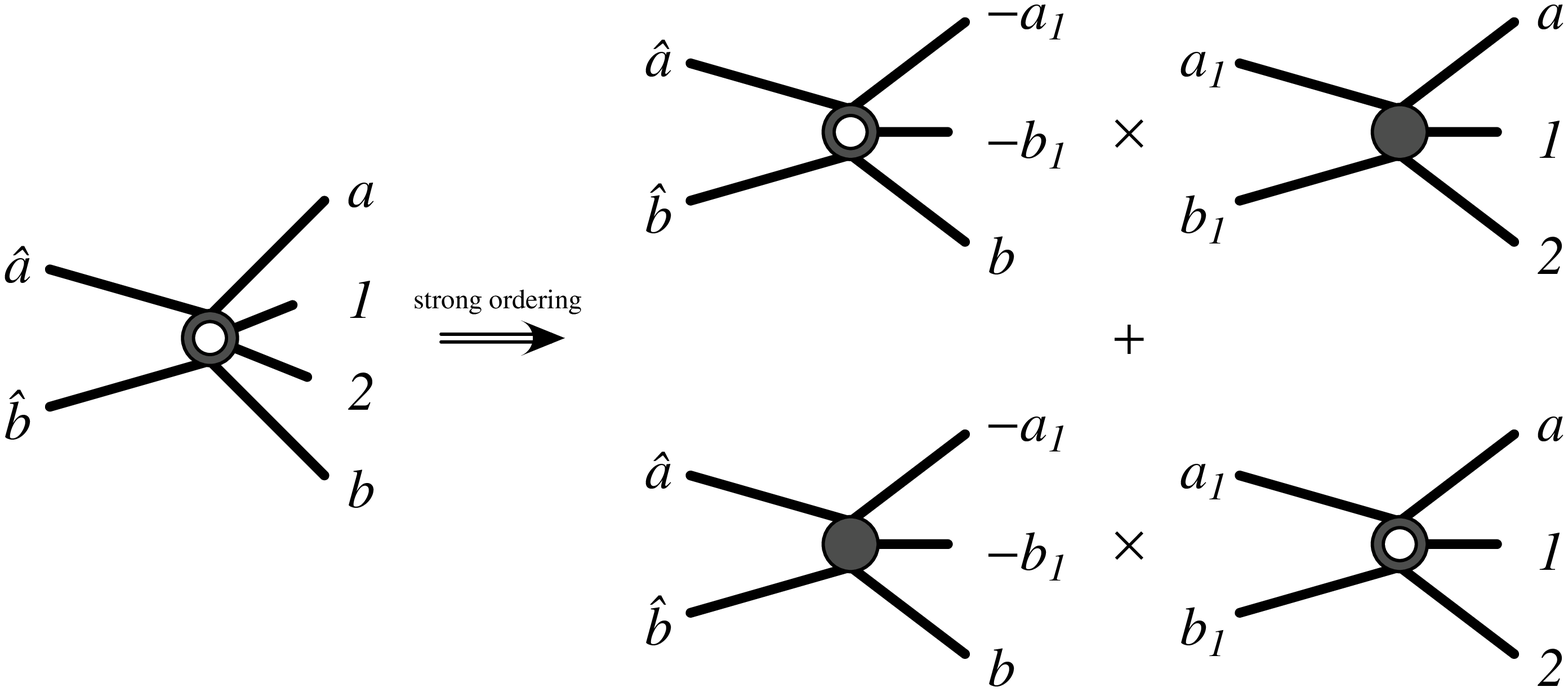}{}\endinsert

Iteratively applying a strong ordering to $m$ singular emissions, 
we obtain the generalization
of eqn.~(\use\MStronglyOrdered),
$$\eqalign{
\Ant^{r\lloop}&(\ah^{\lambda_\ah},\bh^{\lambda_\bh}\leftarrow a,1,\ldots,m,b) 
\inlimit^{\sing{k_1} \ll \sing{k_2}\ll\cdots\ll\sing{k_m} \rightarrow 0}\cr
&\sum_{\phpol \lambda_{\ab1,\bb1,\ldots,\ab{m},\bb{m}}}
\sum_{v_i=0\atop v_1+\cdots+v_m = r}^{r}
\Ant^{v_1\lloop}(\ab1^{\lambda_\ab1},\bb1^{\lambda_\bb1}\leftarrow a,1,2)
\cr &\hphantom{ \sum_{\phpol \lambda_{\ab1,\bb1,\ldots,\ab{m},\bb{m}}} }
\hskip 5mm\times
\Ant^{v_2\lloop}(\ab2^{\lambda_\ab2},\bb2^{\lambda_\bb2}\leftarrow 
     -k_\ab1^{-\lambda_\ab1},-k_\bb1^{-\lambda_\bb1},3)
\cr &\hphantom{ \sum_{\phpol \lambda_{\ab1,\bb1,\ldots,\ab{m},\bb{m}}} }
\hskip 5mm\times\cdots\times
 \Ant^{v_m\lloop}(\ah^{\lambda_\ah},\bh^{\lambda_\bh}\leftarrow 
     -k_\ab{m-1}^{-\lambda_\ab{m-1}},-k_\bb{m-1}^{-\lambda_\bb{m-1}},b)
+\cdots
}\eqn\MStronglyOrderedHigherLoop$$
where the intermediate momenta are defined in eqn.~(\use\NestedMomenta).
This has the same iterated structure as eqn.~(\use\MStronglyOrdered), but with
the total number of loops `distributed' in all possible ways amongst the iterated
antenna amplitudes.

\section{Conclusions}
\vskip 10pt

A detailed understanding of the singular structure of real emission in perturbative
gauge theories is important to the ongoing program of next-to-next-to-leading
order (NNLO) corrections to jet observables.  The antenna functions combine soft
and collinear limits in a simple way, allowing a simpler approach to the 
calculation of integrals over phase space of real-emission.  The strongly-ordered
limits discussed here have two important applications.  Taken to all orders,
along with knowledge of the leading singular structure of virtual corrections,
they allow the resummation of the terms in the matrix elements that give rise
to the leading logarithms for observables with a large ratio of scales.  To NNLO,
they summarize the leading and most singular contributions to real-emission.  
Since they iterate the one-loop emission probabilities, they are considerably simpler
than the full double-singular emission probabilities.  Furthermore, their structure
suggests it should be possible to match the Catani decomposition of two-loop
singularities~[\use\CataniConjecture,\use\StermanTejeda]
 structure directly in the real-emission contributions.

\appendix{Interpolating Functions for Strongly-Ordered Limits}
\vskip 10pt
The reconstruction functions given in 
ref.~[\use\MultipleAntenna] are appropriate only to uniform limits,
where all singular invariants become singular at the same rate.  For
strongly-ordered limits, we should take different forms for $k_{\ah}$
and $k_{\bh}$ in each term of the sum~(\use\MultipleEmissionAntennaDef),
corresponding to different choices of the interpolation functions $r_i$
in each term.  In eq.~(7.8) of ref.~[\use\MultipleAntenna], the following 
form was given for the interpolation functions,
$$
r_j^{\{0\}} \equiv r_j = {k_j\cdot (K_{j\tp1,m}+k_b)\over k_j\cdot K};
\eqn\DoubleRChoice$$
we retain this form for the first and last terms ($\ell=0,m$)
but should instead pick,
$$
r_j^{\{\ell\}} = 1- {k_{\ell+1}\cdot (K_{\ell+1,n}+k_b)\over 
                     k_{\ell+1}\cdot K} (1-r_j^{\{0\}}),
\eqn\NewRChoiceA
$$
for $j\le\ell$ and
$$
r_j^{\{\ell\}} = {k_{\ell}\cdot (k_a+K_{1\cdots\ell})\over 
                     k_{\ell}\cdot K} r_j^{\{0\}},
\eqn\NewRChoiceB
$$
for $j > \ell$.
The reconstructed momenta $\ah^{\{\ell\}}$ and $\bh^{\{\ell\}}$ in
the $\ell$-th term of eq.~(\use\MultipleEmissionAntennaDef) are given
by the same functional forms as given in eq.~(6.2) of
 ref.~[\use\MultipleAntenna], but
with $r_j$ replaced by $r_j^{\{\ell\}}$.

While we cannot eliminate $k_{\ah}$ and $k_{\bh}$ from the antenna amplitudes,
because the phase dependence of their spinor products is needed, 
we can do so in the
squared amplitude.  It is possible to do so without introducing square roots;
indeed, noting that $s_{\ah\bh} = K^2$, we see that $A_1$, given in 
eq.~(10.3) of ref.~[\use\MultipleAntenna], and $E_{1,2}$, given in eq.~(10.6),
are already free of the two reconstructed momenta.  We can put $A_2$ in
such a form as well,
$$\eqalign{
A_2&(\ah,a,1,2,b,\bh) = \cr
&
\frac{4}{s_{a1}\,s_{2b}}\biggl(2 + \frac{4\,s_{12}}{K^2} - \frac{2\,t_{a12}}{K^2} - \frac{5\,t_{12b}}{K^2}\biggr)
+\frac{8}{s_{a1}\,s_{12}}\biggl(3 - \frac{3\,s_{2b}}{K^2} + \frac{4\,{s_{2b}}^2}{(K^2)^2} + \frac{t_{12b}}{K^2} - \frac{5\,s_{2b}\,t_{12b}}{(K^2)^2} + \frac{2\,{t_{12b}}^2}{(K^2)^2}\biggr)
\cr&
-\frac{8}{s_{12}\,s_{2b}}
-\frac{4\,K^2}{s_{a1}\,s_{12}\,s_{2b}}\biggl(\frac{t_{12b}}{K^2} - \frac{t_{a12}\,t_{12b}}{(K^2)^2} - \frac{{t_{12b}}^2}{(K^2)^2}\biggr)
+\frac{8}{s_{12}\,t_{a12}}\biggl(1 + \frac{4\,s_{2b}}{K^2} - \frac{3\,t_{12b}}{K^2} + \frac{{t_{12b}}^2}{(K^2)^2}\biggr)
\cr&
-\frac{8\,K^2}{s_{a1}\,s_{2b}\,t_{a12}}\biggl(2 + \frac{s_{12}}{K^2} + \frac{{s_{12}}^2}{(K^2)^2} - \frac{t_{12b}}{K^2} - \frac{2\,s_{12}\,t_{12b}}{(K^2)^2} + \frac{{t_{12b}}^2}{(K^2)^2}\biggr)
\cr&
+\frac{2}{{s_{12}}^2}\biggl(4 - \frac{12\,s_{2b}}{K^2} + \frac{4\,{s_{2b}}^2}{(K^2)^2} + \frac{8\,t_{12b}}{K^2} - \frac{7\,s_{2b}\,t_{12b}}{(K^2)^2} + \frac{5\,{t_{12b}}^2}{(K^2)^2}\biggr)
-\frac{8}{s_{a1}\,t_{a12}}\biggl(3 + \frac{s_{2b}}{K^2} - \frac{4\,t_{12b}}{K^2}\biggr)
\cr&
+\frac{8\,K^2}{s_{a1}\,s_{12}\,\left( K^2 - s_{2b} - t_{a12} \right) }\biggl(\frac{t_{12b}}{K^2} + \frac{{s_{2b}}^2\,t_{12b}}{(K^2)^3} - \frac{2\,s_{2b}\,{t_{12b}}^2}{(K^2)^3} + \frac{{t_{12b}}^3}{(K^2)^3}\biggr)
\cr&
+\frac{-8\,K^2}{s_{12}\,s_{2b}\,t_{a12}}\biggl(-1 + \frac{{s_{a1}}^3}{(K^2)^3} - \frac{2\,{s_{a1}}^2}{(K^2)^2} + \frac{s_{a1}}{K^2} + \frac{t_{12b}}{K^2} - \frac{{t_{12b}}^2}{(K^2)^2} + \frac{{t_{12b}}^3}{(K^2)^3}\biggr)
-\frac{16\,s_{a1}}{{s_{12}}^2\,t_{a12}}
\cr&
-\frac{16\,K^2}{s_{a1}\,\left( K^2 - s_{2b} - t_{a12} \right) \,t_{a12}}\biggl(\frac{s_{2b}}{K^2} + \frac{2\,{s_{2b}}^2}{(K^2)^2} + \frac{t_{12b}}{K^2} - \frac{4\,s_{2b}\,t_{12b}}{(K^2)^2} + \frac{2\,{t_{12b}}^2}{(K^2)^2}\biggr)
\cr&
-\frac{8\,K^2}{s_{12}\,\left( K^2 - s_{2b} - t_{a12} \right) \,t_{a12}}\biggl(\frac{2\,t_{12b}}{K^2} - \frac{s_{2b}\,{t_{12b}}^2}{(K^2)^3} + \frac{{t_{12b}}^3}{(K^2)^3}\biggr)
+\frac{24}{{t_{a12}}^2}
+\frac{16\,s_{12}}{s_{a1}\,{t_{a12}}^2}
+\frac{16\,s_{a1}}{s_{12}\,{t_{a12}}^2}
\cr&
+\frac{8\,{s_{a1}}^2}{{s_{12}}^2\,{t_{a12}}^2}
-\frac{8\,K^2}{s_{12}\,s_{2b}\,\left( -s_{a1} + K^2 - t_{12b} \right) }\biggl(\frac{2\,{s_{a1}}^2}{(K^2)^2} - \frac{s_{a1}\,t_{a12}}{(K^2)^2}\biggr)
\cr&
-\frac{8\,K^2}{s_{a1}\,s_{12}\,\left( K^2 - t_{a12} - t_{12b} \right) }\biggl(2 - \frac{3\,s_{2b}}{K^2} + \frac{2\,{s_{2b}}^2}{(K^2)^2} - \frac{{s_{2b}}^3}{(K^2)^3}\biggr)
\cr&
-\frac{8\,K^2}{s_{12}\,t_{a12}\,\left( K^2 - t_{a12} - t_{12b} \right) }\biggl(\frac{-2\,s_{2b}}{K^2} + \frac{{s_{2b}}^2}{(K^2)^2} + \frac{t_{12b}}{K^2}\biggr)
+\frac{8\,{s_{a1}}^3}{K^2\,s_{12}\,s_{2b}\,t_{a12}\,\left( -s_{a1} + K^2 - t_{12b} \right) }
\cr&
-\frac{8\,K^2}{s_{a1}\,t_{a12}\,\left( K^2 + s_{12} - t_{a12} - t_{12b} \right) }\biggl(2 - \frac{3\,s_{2b}}{K^2} + \frac{2\,{s_{2b}}^2}{(K^2)^2} - \frac{{s_{2b}}^3}{(K^2)^3}\biggr)
\cr&
+\frac{8\,(K^2)^2}{s_{a1}\,s_{2b}\,t_{a12}\,\left( K^2 + s_{12} - t_{a12} - t_{12b} \right) }
-\frac{8\,K^2}{s_{a1}\,t_{a12}\,t_{12b}}\biggl(2 - \frac{3\,s_{2b}}{K^2} + \frac{2\,{s_{2b}}^2}{(K^2)^2} - \frac{{s_{2b}}^3}{(K^2)^3}\biggr)
\cr &
-\frac{8\,{t_{12b}}^2}{s_{12}\,s_{2b}\,t_{a12}\,\left( -K^2 + t_{a12} + t_{12b} \right) }
+\frac{8\,{s_{2b}}^2}{s_{12}\,t_{a12}\,\left( -K^2 + s_{2b} + t_{a12} \right) \,\left( -K^2 + t_{a12} + t_{12b} \right) }
}\eqn\NewATwo$$
which expression is also valid in the strongly-ordered limits.

\listrefs
\bye

%% file: header.tex
\newbox\SlashedBox 
\def\slashed#1{\setbox\SlashedBox=\hbox{#1}
\hbox to 0pt{\hbox to 1\wd\SlashedBox{\hfil/\hfil}\hss}{#1}}
\def\hboxtosizeof#1#2{\setbox\SlashedBox=\hbox{#1}
\hbox to 1\wd\SlashedBox{#2}}

\def\dfrac#1/#2{%
\hskip-.2em\kern .2em\raise.4ex\hbox{\the\scriptfont0 #1}\kern-.2em/%
\kern -.15em\lower.35ex\hbox{\the\scriptfont0 #2}}
\def\mathslashed#1{\setbox\SlashedBox=\hbox{$#1$}
\hbox to 0pt{\hbox to 1\wd\SlashedBox{\hfil/\hfil}\hss}#1}

\def\clap#1{\hbox to 0pt{\hss#1\hss}}

\def\ifsmall{\iffalse}  
\def\titlepagefont{}  

\def\DefineTeXgraphics{%
\special{ps::[global] /TeXgraphics { } def}}  

\def\today{\ifcase\month\or January\or February\or March\or April\or May
\or June\or July\or August\or September\or October\or November\or
December\fi\space\number\day, \number\year}
\def\eatPrefix19{}
\def\Year{\expandafter\eatPrefix\the\year}
\newcount\hours \newcount\minutes
\def\monthname{\ifcase\month\or
January\or February\or March\or April\or May\or June\or July\or
August\or September\or October\or November\or December\fi}
\def\shortmonthname{\ifcase\month\or
Jan\or Feb\or Mar\or Apr\or May\or Jun\or Jul\or
Aug\or Sep\or Oct\or Nov\or Dec\fi}

\def\TimeStamp{\hours\the\time\divide\hours by60%
\minutes -\the\time\divide\minutes by60\multiply\minutes by60%
\advance\minutes by\the\time%
${\rm \shortmonthname}\cdot\if\day<10{}0\fi\the\day\cdot\the\year%
\qquad\the\hours:\if\minutes<10{}0\fi\the\minutes$}




\def\Title#1{%
\vskip 1in{\titlefont\centerline{#1}}\vskip .5in}
 
\def\Date#1{\leftline{#1}\tenrm\supereject%
\global\hsize=\hsbody\global\hoffset=\hbodyoffset%
\footline={\hss\tenrm\folio\hss}}

\newif\ifdraftmode
\newif\ifleftlabels  

\def\nolabels{\def\wrlabeL##1{}\def\eqlabeL##1{}\def\reflabeL##1{}}
\def\writelabels{\def\wrlabeL##1{\leavevmode\vadjust{\rlap{\smash%
{\line{{\escapechar=` \hfill\rlap{\sevenrm\hskip.03in\string##1}}}}}}}%
\def\eqlabeL##1{{\escapechar-1\rlap{\sevenrm\hskip.05in\string##1}}}%
\def\reflabeL##1{\noexpand\rlap{\noexpand\sevenrm[\string##1]}}}
\def\writeleftlabels{\def\wrlabeL##1{\leavevmode\vadjust{\rlap{\smash%
{\line{{\escapechar=` \hfill\rlap{\sevenrm\hskip.03in\string##1}}}}}}}%
\def\eqlabeL##1{{\escapechar-1%
\rlap{\sixrm\hskip.05in\string##1}%
\llap{\sevenrm\string##1\hskip.03in\hbox to \hsize{}}}}%
\def\reflabeL##1{\noexpand\rlap{\noexpand\sevenrm[\string##1]}}}
\nolabels

\input hyperbasics.tex

\newdimen\fullhsize
\newdimen\hstitle
\hstitle=\hsize 
\newdimen\hsbody
\hsbody=\hsize 
\newdimen\hbodyoffset
\hbodyoffset=\hoffset 
\newbox\leftpage
\def\abstract#1{#1}
\def\rotated{\special{ps: landscape}
\magnification=1000  
\baselineskip=14pt
\global\hstitle=9truein\global\hsbody=4.75truein
\global\vsize=7truein\global\voffset=-.31truein
\global\hoffset=-0.54in\global\hbodyoffset=-.54truein
\global\fullhsize=10truein
\def\DefineTeXgraphics{%
\special{ps::[global] 
/TeXgraphics {currentpoint translate 0.7 0.7 scale
              -80 0.72 mul -1000 0.72 mul translate} def}}
\let\lr=L
\def\ifsmall{\iftrue}
\def\titlepagefont{\twelvepoint}
\trueseventeenpoint
\def\almostshipout##1{\if L\lr \count1=1
      \global\setbox\leftpage=##1 \global\let\lr=R
   \else \count1=2
      \shipout\vbox{\hbox to\fullhsize{\box\leftpage\hfil##1}}
      \global\let\lr=L\fi}

\output={\ifnum\count0=1 
 \shipout\vbox{\hbox to \fullhsize{\hfill\pagebody\hfill}}\advancepageno
 \else
 \almostshipout{\leftline{\vbox{\pagebody\makefootline}}}\advancepageno 
 \fi}

\def\abstract##1{{\leftskip=1.5in\rightskip=1.5in ##1\par}} }

\def\linemessage#1{\immediate\write16{#1}}

\global\newcount\secno \global\secno=0
\global\newcount\appno \global\appno=0
\global\newcount\meqno \global\meqno=1
\global\newcount\subsecno \global\subsecno=0
\global\newcount\figno \global\figno=0

\newif\ifAnyCounterChanged
\let\terminator=\relax
\def\normalize#1{\ifx#1\terminator\let\next=\relax\else%
\if#1i\aftergroup i\else\if#1v\aftergroup v\else\if#1x\aftergroup x%
\else\if#1l\aftergroup l\else\if#1c\aftergroup c\else%
\if#1m\aftergroup m\else%
\if#1I\aftergroup I\else\if#1V\aftergroup V\else\if#1X\aftergroup X%
\else\if#1L\aftergroup L\else\if#1C\aftergroup C\else%
\if#1M\aftergroup M\else\aftergroup#1\fi\fi\fi\fi\fi\fi\fi\fi\fi\fi\fi\fi%
\let\next=\normalize\fi%
\next}
\def\makeNormal#1#2{\def\doNormalDef{\edef#1}\begingroup%
\aftergroup\doNormalDef\aftergroup{\normalize#2\terminator\aftergroup}%
\endgroup}

\def\warnIfChanged#1#2{%
\ifundef#1
\else\begingroup%
\edef\oldDefinitionOfCounter{#1}\edef\newDefinitionOfCounter{#2}%
\ifx\oldDefinitionOfCounter\newDefinitionOfCounter%
\else%
\linemessage{Warning: definition of \noexpand#1 has changed.}%
\global\AnyCounterChangedtrue\fi\endgroup\fi}

\def\Section#1{\global\advance\secno by1\relax\global\meqno=1%
\global\subsecno=0%
\bigbreak\bigskip
\centerline{\twelvepoint \bf %
\the\secno. #1}%
\par\nobreak\medskip\nobreak}
\def\tagsection#1{%
\warnIfChanged#1{\the\secno}%
\xdef#1{\the\secno}%
\ifWritingAuxFile\immediate\write\auxfile{\noexpand\xdef\noexpand#1{#1}}\fi%
}
\def\section{\Section}
\def\Subsection#1{\global\advance\subsecno by1\relax\medskip %
\leftline{\bf\the\secno.\the\subsecno\ #1}%
\par\nobreak\smallskip\nobreak}
\def\tagsubsection#1{%
\warnIfChanged#1{\the\secno.\the\subsecno}%
\xdef#1{\the\secno.\the\subsecno}%
\ifWritingAuxFile\immediate\write\auxfile{\noexpand\xdef\noexpand#1{#1}}\fi%
}

\def\subsection{\Subsection}

\def\romappno{\uppercase\expandafter{\romannumeral\appno}}
\def\makeNormalizedRomappno{%
\expandafter\makeNormal\expandafter\normalizedromappno%
\expandafter{\romannumeral\appno}%
\edef\normalizedromappno{\uppercase{\normalizedromappno}}}
\def\Appendix#1{\global\advance\appno by1\relax\global\meqno=1\global\secno=0%
\global\subsecno=0%
\bigbreak\bigskip
\centerline{\twelvepoint \bf Appendix %
\romappno. #1}%
\par\nobreak\medskip\nobreak}
\def\tagappendix#1{\makeNormalizedRomappno%
\warnIfChanged#1{\normalizedromappno}%
\xdef#1{\normalizedromappno}%
\ifWritingAuxFile\immediate\write\auxfile{\noexpand\xdef\noexpand#1{#1}}\fi%
}
\def\appendix{\Appendix}
\def\Subappendix#1{\global\advance\subsecno by1\relax\medskip %
\leftline{\bf\romappno.\the\subsecno\ #1}%
\par\nobreak\smallskip\nobreak}
\def\tagsubappendix#1{\makeNormalizedRomappno%
\warnIfChanged#1{\normalizedromappno.\the\subsecno}%
\xdef#1{\normalizedromappno.\the\subsecno}%
\ifWritingAuxFile\immediate\write\auxfile{\noexpand\xdef\noexpand#1{#1}}\fi%
}

\def\eqn#1{\makeNormalizedRomappno%
\ifnum\secno>0%
  \warnIfChanged#1{\the\secno.\the\meqno}%
  \eqno(\the\secno.\the\meqno)\xdef#1{\the\secno.\the\meqno}%
     \global\advance\meqno by1
\else\ifnum\appno>0%
  \warnIfChanged#1{\normalizedromappno.\the\meqno}%
  \eqno({\rm\romappno}.\the\meqno)%
      \xdef#1{\normalizedromappno.\the\meqno}%
     \global\advance\meqno by1
\else%
  \warnIfChanged#1{\the\meqno}%
  \eqno(\the\meqno)\xdef#1{\the\meqno}%
     \global\advance\meqno by1
\fi\fi%
\eqlabeL#1%
\ifWritingAuxFile\immediate\write\auxfile{\noexpand\xdef\noexpand#1{#1}}\fi%
}
\def\defeqn#1{\makeNormalizedRomappno%
\ifnum\secno>0%
  \warnIfChanged#1{\the\secno.\the\meqno}%
  \xdef#1{\the\secno.\the\meqno}%
     \global\advance\meqno by1
\else\ifnum\appno>0%
  \warnIfChanged#1{\normalizedromappno.\the\meqno}%
  \xdef#1{\normalizedromappno.\the\meqno}%
     \global\advance\meqno by1
\else%
  \warnIfChanged#1{\the\meqno}%
  \xdef#1{\the\meqno}%
     \global\advance\meqno by1
\fi\fi%
\eqlabeL#1%
\ifWritingAuxFile\immediate\write\auxfile{\noexpand\xdef\noexpand#1{#1}}\fi%
}
\def\anoneqn{\makeNormalizedRomappno%
\ifnum\secno>0
  \eqno(\the\secno.\the\meqno)%
     \global\advance\meqno by1
\else\ifnum\appno>0
  \eqno({\rm\normalizedromappno}.\the\meqno)%
     \global\advance\meqno by1
\else
  \eqno(\the\meqno)%
     \global\advance\meqno by1
\fi\fi%
}
\def\mfig#1#2{\ifx#20
\else\global\advance\figno by1%
\relax#1\the\figno%
\warnIfChanged#2{\the\figno}%
\xdef#2{\the\figno}%
\reflabeL#2%
\ifWritingAuxFile\immediate\write\auxfile{\noexpand\xdef\noexpand#2{#2}}\fi\fi%
}

\def\fig#1{\mfig{fig.\ }#1}

\catcode`@=11 

\newif\ifFiguresInText\FiguresInTexttrue
\newif\if@FigureFileCreated
\newwrite\capfile
\newwrite\figfile

\newif\ifcaption
\captiontrue
\def\captionsize{\tenrm}
\def\PlaceTextFigure#1#2#3#4{%
\vskip 0.5truein%
\noindent#3\hfil\epsfbox{#4}\hfil\break%
\ifcaption\vskip 5pt\noindent\hfil\vbox{\captionsize \noindent Figure #1. #2}\hfil\fi%
\vskip10pt}
\def\PlaceEndFigure#1#2{%
\epsfxsize=\hsize\epsfbox{#2}\vfill\centerline{Figure #1.}\eject}

\def\LoadFigure#1#2#3#4{%
\vphantom{\mfig{}#1}
\ifx#10
\else
\fi
\ifFiguresInText
\PlaceTextFigure{#1}{#2}{#3}{#4}%
\else
\if@FigureFileCreated\else%
\immediate\openout\capfile=\jobname.caps%
\immediate\openout\figfile=\jobname.figs%
@FigureFileCreatedtrue\fi%
\immediate\write\capfile{\noexpand\item{Figure \noexpand#1.\ }{#2}\vskip10pt}%
\immed	iate\write\figfile{\noexpand\PlaceEndFigure\noexpand#1{\noexpand#4}}%
\fi}

\def\listfigs{\ifFiguresInText\else%
\vfill\eject\immediate\closeout\capfile
\immediate\closeout\figfile%
\centerline{{\bf Figures}}\bigskip\frenchspacing%
\catcode`@=11 
\def\captionsize{\tenrm}
\input \jobname.caps\vfill\eject\nonfrenchspacing%
\catcode`\@=\active
\catcode`@=12  
\input\jobname.figs\fi}

\font\ninerm=cmr9
\font\eightrm=cmr8
\font\sixrm=cmr6

\def\loadtrueseventeenpoint{
 \font\seventeenrm=cmr10 at 17.28truept
 \font\seventeeni=cmmi10 at 17.28truept
 \font\seventeenbf=cmbx10 at 17.28truept
 \font\seventeenit=cmti10 at 17.28truept
 \font\seventeensl=cmsl10 at 17.28truept
 \font\seventeensy=cmsy10 at 17.28truept
}
\def\loadfourteenpoint{
\font\fourteenrm=cmr10 at 14.4pt
\font\fourteeni=cmmi10 at 14.4pt
\font\fourteenit=cmti10 at 14.4pt
\font\fourteensl=cmsl10 at 14.4pt
\font\fourteensy=cmsy10 at 14.4pt
\font\fourteenbf=cmbx10 at 14.4pt
}
\def\loadtruetwelvepoint{
\font\twelverm=cmr10 at 12truept
\font\twelvei=cmmi10 at 12truept
\font\twelveit=cmti10 at 12truept
\font\twelvesl=cmsl10 at 12truept
\font\twelvesy=cmsy10 at 12truept
\font\twelvebf=cmbx10 at 12truept
\font\twelvesc=cmcsc10 at 12truept
}

\font\ninei=cmmi9
\font\eighti=cmmi8
\font\sixi=cmmi6
\skewchar\ninei='177 \skewchar\eighti='177 \skewchar\sixi='177

\font\ninesy=cmsy9
\font\eightsy=cmsy8
\font\sixsy=cmsy6
\skewchar\ninesy='60 \skewchar\eightsy='60 \skewchar\sixsy='60

\font\ninebf=cmbx9
\font\eightbf=cmbx8
\font\sixbf=cmbx6

\font\ninett=cmtt9
\font\eighttt=cmtt8

\hyphenchar\tentt=-1 
\hyphenchar\ninett=-1
\hyphenchar\eighttt=-1         

\font\ninesl=cmsl9
\font\eightsl=cmsl8

\font\nineit=cmti9
\font\eightit=cmti8
\font\sevenit=cmti7

\scriptfont\itfam=\sevenit


                      
\newskip\ttglue
\def\tenpoint{\def\rm{\fam0\tenrm}%
  \textfont0=\tenrm \scriptfont0=\sevenrm \scriptscriptfont0=\fiverm
  \textfont1=\teni \scriptfont1=\seveni \scriptscriptfont1=\fivei
  \textfont2=\tensy \scriptfont2=\sevensy \scriptscriptfont2=\fivesy
  \textfont3=\tenex \scriptfont3=\tenex \scriptscriptfont3=\tenex
  \def\it{\fam\itfam\tenit}%
      \textfont\itfam=\tenit\scriptfont\itfam=\sevenit
  \def\sl{\fam\slfam\tensl}\textfont\slfam=\tensl
  \def\bf{\fam\bffam\tenbf}\textfont\bffam=\tenbf \scriptfont\bffam=\sevenbf
  \scriptscriptfont\bffam=\fivebf
  \normalbaselineskip=12pt
  \let\sc=\eightrm
  \let\big=\tenbig
  \setbox\strutbox=\hbox{\vrule height8.5pt depth3.5pt width\z@}%
  \normalbaselines\rm}

\def\twelvepoint{\def\rm{\fam0\twelverm}%
  \textfont0=\twelverm \scriptfont0=\ninerm \scriptscriptfont0=\sevenrm
  \textfont1=\twelvei \scriptfont1=\ninei \scriptscriptfont1=\seveni
  \textfont2=\twelvesy \scriptfont2=\ninesy \scriptscriptfont2=\sevensy
  \textfont3=\tenex \scriptfont3=\tenex \scriptscriptfont3=\tenex
  \def\it{\fam\itfam\twelveit}\textfont\itfam=\twelveit
  \def\sl{\fam\slfam\twelvesl}\textfont\slfam=\twelvesl
  \def\bf{\fam\bffam\twelvebf}\textfont\bffam=\twelvebf%
  \scriptfont\bffam=\ninebf
  \scriptscriptfont\bffam=\sevenbf
  \normalbaselineskip=12pt
  \let\sc=\eightrm
  \let\big=\tenbig
  \setbox\strutbox=\hbox{\vrule height8.5pt depth3.5pt width\z@}%
  \normalbaselines\rm}

\def\fourteenpoint{\def\rm{\fam0\fourteenrm}%
  \textfont0=\fourteenrm \scriptfont0=\tenrm \scriptscriptfont0=\sevenrm
  \textfont1=\fourteeni \scriptfont1=\teni \scriptscriptfont1=\seveni
  \textfont2=\fourteensy \scriptfont2=\tensy \scriptscriptfont2=\sevensy
  \textfont3=\tenex \scriptfont3=\tenex \scriptscriptfont3=\tenex
  \def\it{\fam\itfam\fourteenit}\textfont\itfam=\fourteenit
  \def\sl{\fam\slfam\fourteensl}\textfont\slfam=\fourteensl
  \def\bf{\fam\bffam\fourteenbf}\textfont\bffam=\fourteenbf%
  \scriptfont\bffam=\tenbf
  \scriptscriptfont\bffam=\sevenbf
  \normalbaselineskip=17pt
  \let\sc=\elevenrm
  \let\big=\tenbig                                          
  \setbox\strutbox=\hbox{\vrule height8.5pt depth3.5pt width\z@}%
  \normalbaselines\rm}

\def\seventeenpoint{\def\rm{\fam0\seventeenrm}%
  \textfont0=\seventeenrm \scriptfont0=\fourteenrm \scriptscriptfont0=\tenrm
  \textfont1=\seventeeni \scriptfont1=\fourteeni \scriptscriptfont1=\teni
  \textfont2=\seventeensy \scriptfont2=\fourteensy \scriptscriptfont2=\tensy
  \textfont3=\tenex \scriptfont3=\tenex \scriptscriptfont3=\tenex
  \def\it{\fam\itfam\seventeenit}\textfont\itfam=\seventeenit
  \def\sl{\fam\slfam\seventeensl}\textfont\slfam=\seventeensl
  \def\bf{\fam\bffam\seventeenbf}\textfont\bffam=\seventeenbf%
  \scriptfont\bffam=\fourteenbf
  \scriptscriptfont\bffam=\twelvebf
  \normalbaselineskip=21pt
  \let\sc=\fourteenrm
  \let\big=\tenbig                                          
  \setbox\strutbox=\hbox{\vrule height 12pt depth 6pt width\z@}%
  \normalbaselines\rm}

\def\ninepoint{\def\rm{\fam0\ninerm}%
  \textfont0=\ninerm \scriptfont0=\sixrm \scriptscriptfont0=\fiverm
  \textfont1=\ninei \scriptfont1=\sixi \scriptscriptfont1=\fivei
  \textfont2=\ninesy \scriptfont2=\sixsy \scriptscriptfont2=\fivesy
  \textfont3=\tenex \scriptfont3=\tenex \scriptscriptfont3=\tenex
  \def\it{\fam\itfam\nineit}\textfont\itfam=\nineit
  \def\sl{\fam\slfam\ninesl}\textfont\slfam=\ninesl
  \def\bf{\fam\bffam\ninebf}\textfont\bffam=\ninebf \scriptfont\bffam=\sixbf
  \scriptscriptfont\bffam=\fivebf
  \normalbaselineskip=11pt
  \let\sc=\sevenrm
  \let\big=\ninebig
  \setbox\strutbox=\hbox{\vrule height8pt depth3pt width\z@}%
  \normalbaselines\rm}

\def\eightpoint{\def\rm{\fam0\eightrm}%
  \textfont0=\eightrm \scriptfont0=\sixrm \scriptscriptfont0=\fiverm%
  \textfont1=\eighti \scriptfont1=\sixi \scriptscriptfont1=\fivei%
  \textfont2=\eightsy \scriptfont2=\sixsy \scriptscriptfont2=\fivesy%
  \textfont3=\tenex \scriptfont3=\tenex \scriptscriptfont3=\tenex%
  \def\it{\fam\itfam\eightit}\textfont\itfam=\eightit%
  \def\sl{\fam\slfam\eightsl}\textfont\slfam=\eightsl%
  \def\bf{\fam\bffam\eightbf}\textfont\bffam=\eightbf \scriptfont\bffam=\sixbf%
  \scriptscriptfont\bffam=\fivebf%
  \normalbaselineskip=9pt%
  \let\sc=\sixrm%
  \let\big=\eightbig%
  \setbox\strutbox=\hbox{\vrule height7pt depth2pt width\z@}%
  \normalbaselines\rm}
  \let\sc=\eightrm

\def\tenbig#1{{\hbox{$\left#1\vbox to8.5pt{}\right.\n@space$}}}
\def\ninebig#1{{\hbox{$\textfont0=\tenrm\textfont2=\tensy
  \left#1\vbox to7.25pt{}\right.\n@space$}}}
\def\eightbig#1{{\hbox{$\textfont0=\ninerm\textfont2=\ninesy
  \left#1\vbox to6.5pt{}\right.\n@space$}}}

\def\footnote#1{\edef\@sf{\spacefactor\the\spacefactor}#1\@sf
      \insert\footins\bgroup\eightpoint
      \interlinepenalty100 \let\par=\endgraf
        \leftskip=\z@skip \rightskip=\z@skip
        \splittopskip=10pt plus 1pt minus 1pt \floatingpenalty=20000
        \smallskip\item{#1}\bgroup\strut\aftergroup\@foot\let\next}
\skip\footins=12pt plus 2pt minus 4pt 
\dimen\footins=30pc 

\newinsert\margin
\dimen\margin=\maxdimen
\def\titlefont{\seventeenpoint}
\loadtruetwelvepoint 
\loadtrueseventeenpoint

\def\eatOne#1{}
\def\ifundef#1{\expandafter\ifx%
\csname\expandafter\eatOne\string#1\endcsname\relax}
\def\notTrue{\iffalse}\def\isTrue{\iftrue}
\def\ifdef#1{{\ifundef#1%
\aftergroup\notTrue\else\aftergroup\isTrue\fi}}
\def\use#1{\ifundef#1\linemessage{Warning: \string#1 is undefined.}%
{\tt \string#1}\else#1\fi}



%
\catcode`"=11
\let\quote="
\catcode`"=12
\chardef\foo="22
\global\newcount\refno \global\refno=1
\newwrite\rfile
\newlinechar=`\^^J
\def\@ref#1#2{\the\refno\n@ref#1{#2}}
\def\h@ref#1#2#3{\href{#3}{\the\refno}\n@ref#1{#2}}
\def\n@ref#1#2{\xdef#1{\the\refno}%
\ifnum\refno=1\immediate\openout\rfile=\jobname.refs\fi%
\immediate\write\rfile{\noexpand\item{[\noexpand#1]\ }#2.}%
\global\advance\refno by1}
\def\nref{\n@ref} 
\def\ref{\@ref}   
\def\hrref{\h@ref}
\def\lref#1#2{\the\refno\xdef#1{\the\refno}%
\ifnum\refno=1\immediate\openout\rfile=\jobname.refs\fi%
\immediate\write\rfile{\noexpand\item{[\noexpand#1]\ }#2\semi}%
\global\advance\refno by1}
\def\cref#1{\immediate\write\rfile{#1\semi}}

\def\preref#1#2{\gdef#1{\@ref#1{#2}}}

\def\semi{;\hfil\noexpand\break}

\def\listrefs{\vfill\eject\immediate\closeout\rfile
\centerline{{\bf References}}\bigskip\frenchspacing%
\input \jobname.refs\vfill\eject\nonfrenchspacing}

\def\inputAuxIfPresent#1{\immediate\openin1=#1
\ifeof1\message{No file \auxfileName; I'll create one.
}\else\closein1\relax\input\auxfileName\fi%
}
\def\NPB{Nucl.\ Phys.\ B}




\newif\ifWritingAuxFile
\newwrite\auxfile
\def\SetUpAuxFile{%
\xdef\auxfileName{\jobname.aux}%
\inputAuxIfPresent{\auxfileName}%
\WritingAuxFiletrue%
\immediate\openout\auxfile=\auxfileName}

\def\L{\left(}\def\R{\right)}
\def\LP{\left.}
\def\LB{\left[}\def\RB{\right]}

\def\LV{\left|}\def\RV{\right|}

\def\bye{\par\vfill\supereject%
\ifAnyCounterChanged\linemessage{
Some counters have changed.  Re-run tex to fix them up.}\fi%
\end}

\catcode`\@=\active
\catcode`@=12  
\catcode`\"=\active

%% file: gaugedefs.tex
\def\L{\left(}
\def\R{\right)}

\def\Tr{\mathop{\rm Tr}\nolimits}
\def\Gr{\mathop{\rm Gr}\nolimits}

\def\LP{\left.}

\def\dl^#1_#2{\delta^{#1}{}_{#2}}

\catcode`@=11  
\def\meqalign#1{\,\vcenter{\openup1\jot\m@th
   \ialign{\strut\hfil$\displaystyle{##}$ && $\displaystyle{{}##}$\hfil
             \crcr#1\crcr}}\,}
\catcode`@=12  


\baselineskip 15pt
\overfullrule 0.5pt

%% file: spinordef.tex
\def\Tr{\mathop{\rm Tr}\nolimits}

\def\L{\left(}\def\R{\right)}
\def\LP{\left.}
\def\spa#1.#2{\left\langle#1\,#2\right\rangle}
\def\spb#1.#2{\left[#1\,#2\right]}
\def\lor#1.#2{\left(#1\,#2\right)}
\def\sand#1.#2.#3{%
\left\langle\smash{#1}{\vphantom1}^{-}\right|{#2}%
\left|\smash{#3}{\vphantom1}^{-}\right\rangle}
\def\sandp#1.#2.#3{%
\left\langle\smash{#1}{\vphantom1}^{-}\right|{#2}%
\left|\smash{#3}{\vphantom1}^{+}\right\rangle}
\def\sandpp#1.#2.#3{%
\left\langle\smash{#1}{\vphantom1}^{+}\right|{#2}%
\left|\smash{#3}{\vphantom1}^{+}\right\rangle}
\def\sandpm#1.#2.#3{%
\left\langle\smash{#1}{\vphantom1}^{+}\right|{#2}%
\left|\smash{#3}{\vphantom1}^{-}\right\rangle}
\def\sandmp#1.#2.#3{%
   \left\langle\smash{#1}{\vphantom1}^{-}\right|{#2}%
    \left|\smash{#3}{\vphantom1}^{+}\right\rangle}
\catcode`@=11  
\def\meqalign#1{\,\vcenter{\openup1\jot\m@th
   \ialign{\strut\hfil$\displaystyle{##}$ && $\displaystyle{{}##}$\hfil
             \crcr#1\crcr}}\,}
\catcode`@=12  